\documentclass[12pt]{article}
\usepackage{a4wide,amssymb,cite}
\pdfoutput=1

\usepackage{graphicx}
\usepackage{a4wide,amssymb,graphicx}
\usepackage{epsfig}
\usepackage[usenames,dvipsnames]{color}
\usepackage{slashed}
\usepackage{amssymb,cite,graphicx}
\usepackage{slashed}
\usepackage{amsmath,bm,bbm}
\usepackage{amsfonts}
\usepackage[titletoc,title]{appendix}
\usepackage[small]{caption}
\usepackage[margin=1in]{geometry}
\usepackage[multiple]{footmisc}
\usepackage{mathtools}
\usepackage{slashed}
\usepackage[nottoc]{tocbibind}
\usepackage{xcolor}

\usepackage{tikz}
\usetikzlibrary{shapes,arrows,positioning,automata,backgrounds,calc,er,patterns}
\usepackage[compat=1.1.0]{tikz-feynman}
\usetikzlibrary{trees}
\usetikzlibrary{decorations.pathmorphing}
\usetikzlibrary{decorations.markings}

\newcommand{\be}{\begin{equation}}
\newcommand{\ee}{\end{equation}}
\newcommand{\bea}{\begin{eqnarray}}
\newcommand{\eea}{\end{eqnarray}}

\def\circa#1{\,\raise.3ex\hbox{$#1$\kern-.75em\lower1ex\hbox{$\sim$}}\,}

\begin{document}

\begin{titlepage}
%
%


%

\begin{centering}
\vspace{1cm}
{\Large {\bf The $SU(2)_D$ lepton portals \vspace{.2cm} \\ for muon $g-2$, $W$ boson mass and dark matter}} \\

\vspace{1.5cm}

{\bf Seong-Sik Kim$^\ddagger$, Hyun Min Lee$^\dagger$, Adriana G. Menkara$^\sharp$, \\ and Kimiko Yamashita$^\star$ }
\vspace{.5cm}

{\it  Department of Physics, Chung-Ang University, Seoul 06974, Korea.}

\vspace{.5cm}


\end{centering}
\vspace{2cm}

\begin{abstract}
\noindent
We propose a novel model for lepton flavor and dark matter based on the $SU(2)_D$ gauge symmetry and vector-like leptons in its fundamental representations. We introduce a dark $SU(2)_D$ Higgs doublet and a Higgs bi-doublet for the mass mixing between the vector-like lepton and the lepton. As a result, the seesaw lepton masses are generated and there are sizable one-loop contributions to the muon $g-2$ via the $SU(2)_D$ gauge bosons and the relatively heavy vector-like lepton, as indicated in Fermilab E989. The tree-level mass mixing between the $Z$ boson and the isospin neutral gauge boson of $SU(2)_D$ in our model accounts for the shift in the $W$ boson mass, being consistent with Tevatron CDFII. Finally, we show that the isospin charged gauge boson of $SU(2)_D$ becomes a plausible candidate for dark matter with a small mass splitting tied up to the modified $W$ boson mass, and there is a viable parameter space where the favored corrections to the muon $g-2$ and the $W$ boson mass and  the dark matter constraints are simultaneously fulfilled.

\end{abstract}

\vspace{3cm}

\begin{flushleft} 
$^\ddagger$Email: sskim.working@gmail.com \\
$^\dagger$Email: hminlee@cau.ac.kr \\
$^\sharp$Email: amenkara@cau.ac.kr \\
$^\star$Email: kimikoy@cau.ac.kr
\end{flushleft}

\end{titlepage}


\section{Introduction}

There are a number of unsolved problems in the Standard Model (SM), among which the flavor structure of the matter sector and the origin of dark matter have motivated a lot of interesting ideas beyond the SM. Although there has been no direct hint for new particles or symmetries at a TeV scale or so yet, it is a nontrivial but important task to test the models for flavor and dark matter in general. 

Flavor Changing Neutral Currents and tests of lepton flavor universality can provide interesting ways to probe the inner structure of nature beyond the SM.  Moreover, the electric dipole moment and magnetic dipole moment of leptons and the electroweak precision observables are vulnerable to the presence of heavy particles not far from the electroweak scale appearing in radiative corrections. 
Thus, it is plausible that some of new particles underlying in the flavor structure might be regarded as dark matter candidates, in a similar spirit that low-energy supersymmetry introduced to solve the hierarchy problem leads to neutral superpartners  playing a role of dark matter candidates.
 
There has been a longstanding anomaly in the muon $g-2$ from Brookhaven E821 \cite{Muong-2:2006rrc}, showing the deviation from the SM prediction \cite{Aoyama:2020ynm}, and it has been recently confirmed by Fermilab E989 \cite{Muong-2:2021ojo} with the combined significance at the level of $4.2\sigma$. Furthermore, the measured $W$ boson mass at Tevatron CDFII \cite{CDF:2022hxs} is deviated with high significance from the SM prediction \cite{Haller:2018nnx}  and the previously measured values \cite{ParticleDataGroup:2020ssz}, stirring renewed interests in the electroweak precision data \cite{fits}, models with extra Higgs multiplets \cite{kimiko,extraH}, and extra fermions or gauge bosons \cite{kimiko,extrastates}. There is a need of improvements on the experimental errors and the theory predictions for those most well measured and understood observables.
Nonetheless, it is also important to make a connection between the muon $g-2$ and the $W$ boson mass in a unified framework \cite{kimiko,unified}, provide more information for further developments and suggest new ideas for other experimental searches such as dark matter experiments. 

In this article, we propose a consistent framework for explaining the muon $g-2$ and the $W$ boson mass as well as the dark matter puzzles simultaneously. We aim to address some flavor puzzles such as small fermion masses, in particular, lepton masses, in the presence of vector-like leptons \cite{seesaw,kimiko}.  In the current work, we introduce an extra $SU(2)_D$ gauge symmetry and vector-like new fermions in its fundamental representation beyond the SM.  Thus, we dub this possibility ``$SU(2)_D$  lepton portals''. 
 Assuming that the SM particles are neutral under the $SU(2)_D$, we find it necessary to introduce an $SU(2)_D$ Higgs doublet and a Higgs bi-doublet under $SU(2)_D$ and $SU(2)_L$ to give the mixing masses between the vector-like lepton and the lepton.
Only after a simultaneous breaking of $SU(2)_D$ and electroweak gauge symmetry, a small seesaw mass for the lepton can be generated. 

As a result, for the vector-like lepton with mass of about TeV or higher scale, we consider the contributions of the vector-like lepton and the $SU(2)_D$ gauge bosons to the muon $g-2$. Moreover, when there is an unbroken $Z_2$ parity, which is originated from a combination of the dark isospin symmetry and a global $U(1)_G$ symmetry,  we regard the isospin-charged gauge bosons ($V^\pm$) of $SU(2)_D$ as being dark matter candidates. 
On the other hand, there is a mass mixing between the isospin-neutral gauge boson ($V^0$) of $SU(2)_D$ and the $Z$ boson, which in turn shifts the $Z$ boson mass to a smaller value at the tree level as compared to the SM prediction.  We also correlate the mass splitting for $V^\pm$ and $V^0$ with the correction to the $\rho$ parameter and show the implications of the parameter space for the muon $g-2$ and the $W$ boson mass towards dark matter constraints, namely, the relic density and the direct detection bounds.

The paper is organized as follows.
We begin with an introduction of the model with $SU(2)_D$ gauge symmetry and vector-like leptons as well as the extended Higgs sector. Then, we show the mass spectra for fermions and bosons and new gauge and Yukawa interactions for them in the presence of the $SU(2)_D$ gauge symmetry. Next we discuss new contributions to the muon $g-2$ and the correction to the $W$ boson mass and look for the consistent parameter space for them. We continue to take $V^\pm$ to be dark matter candidates and consider the conditions from the correct relic density for dark matter and the direct detection bounds. Then, conclusions are drawn.
There is one appendix dealing with the mass matrices for scalar and gauge bosons and the rotation matrices.

\section{The model}

We consider an extension of the SM with an extra local $SU(2)_D$ symmetry.
We introduce an $SU(2)_D$ doublet vector-like lepton $\Psi$ with hypercharge $-1$,  an $SU(2)_D$ doublet Higgs $\Phi_D$ and a Higgs bi-doublet $H'$. We note that under $SU(2)_D\times SU(2)_L$,  $\Phi_D$ and $\Psi$ transform as $\Phi_D\to U_D \Phi_D$ and $\Psi\to U_D \Psi$ whereas the bi-doublet transforms as $H'\to U_L H' U^\dagger_D$. Here, $U_D, U_L$ are the transformation matrices for $SU(2)_D$ and $SU(2)_L$.  We assume that the SM Higgs doublet $H$ and the SM fermions as well as three right-handed neutrinos $\nu_R$ are neutral under the $SU(2)_D$. 

There are similar models in the literature, based on the $SU(2)_D$ symmetry, but mostly for dark matter models without extra charged fermions \cite{su2d,su2d2} and with vector-like quarks \cite{VQportal,VQportal2}. There is a  model construction for the muon $g-2$ with an $SU(2)$ symmetry under which the SM leptons transform nontrivially \cite{su2sm}.

We also impose the $Z_2$ parity on the SM and new fields.  The $Z_2$ parity is identified with $Z_2=e^{i\pi (G+I^D_3)}$ where $G$ is the global $U(1)_G$ charge\footnote{Extra fields beyond the SM transform under the $U(1)_G$ by $\Phi_D\to e^{i\alpha/2}\Phi_D$, $\Psi\to e^{i\alpha/2}\Psi$, $H'\to e^{-i\alpha/2} H'$ and $V\to V$, whereas the SM fields are neutral.} and $I^D_3$ is the dark isospin, which is part of $SU(2)_D$.  As a consequence, if the $Z_2$ parity survives the $SU(2)_D$ symmetry breaking,  the dark isospin-charged gauge bosons, $V^\pm$, can be dark matter candidates, as the lightest $Z_2$ odd particles \cite{VQportal,VQportal2}.
 The representations under $SU(2)_D\times SU(2)_L\times U(1)_Y$ and $Z_2$ parity are given in Table~1.

\begin{table}[hbt!]
  \begin{center}
\scalebox{0.9}{
    \begin{tabular}{c|cccccc}
      \hline\hline 
      & $q_L={\scriptsize\left(\begin{array}{c}u_L\\ d_L \end{array}\right)}$ & $u_{R}$  &  $d_{R}$ & $l_{L}={\scriptsize\left(\begin{array}{c}\nu_L\\ e_L \end{array}\right)}$  & $e_{R}$
      & $H={\scriptsize\left(\begin{array}{c}\phi^+_1\\ \phi^0_1 \end{array}\right)}$  \\
      \hline
      $SU(2)_D\times G_{\rm EW}$ & $(1,2)_{+\frac{1}{6}}$ & $(1,1)_{+\frac{2}{3}}$ & $(1,1)_{-\frac{1}{3}}$
            & $(1,2)_{-\frac{1}{2}}$ & $(1,1)_{-1}$ & $(1,2)_{+\frac{1}{2}}$  \\
                      \hline
            $Z_2$ & $+$ & $+$ & $+$
            & $+$ & $+$ & $+$  \\ 
      \hline\hline
    \end{tabular}} \\[4mm]  
   \scalebox{0.9}{
      \begin{tabular}{c|ccccc}
      \hline\hline
     & $\nu_R$ & $H'={\scriptsize\left(\begin{array}{cc} {\hat \phi}^+_2 & \phi_2^+ \\ {\hat\phi}^0_2 & \phi^0_2 \end{array}\right)}$ & $\Psi={\scriptsize\left(\begin{array}{c}E' \\ E\end{array}\right)}$ & $\Phi_D={\scriptsize\left(\begin{array}{c}\varphi_1\\ \varphi_2\end{array}\right)}$ & $V={\scriptsize\left(\begin{array}{c} V^+ \\ V^0 \\ V^- \end{array}\right)}$ \\
      \hline
      $SU(2)_D\times G_{\rm EW}$ & $(1,1)_0$ & $(2,2)_{+\frac{1}{2}}$ & $(2,1)_{-1}$ & $(2,1)_0$  & $(3,1)_0$ \\
      \hline
      $Z_2$ & $+$ &${\scriptsize\left(\begin{array}{cc} - & + \\ - & + \end{array}\right)}$ & ${\scriptsize\left(\begin{array}{c}- \\ +\end{array}\right)}$ & ${\scriptsize\left(\begin{array}{c}-\\ +\end{array}\right)}$ & ${\scriptsize\left(\begin{array}{c} - \\ + \\ - \end{array}\right)}$ \\
      \hline\hline
    \end{tabular}}
  \end{center}
    \caption{Representations under $ SU(2)_D\times G_{\rm EW}$ with $G_{\rm EW}=SU(2)_L\times U(1)_Y$ and $Z_2$ parities. Here, we denote the $U(1)_Y$ hypercharge in the subscript, for instance, $(1,2)_{+\frac{1}{6}}$, for a quark doublet $q_L$, etc. And the $Z_2$ parities shown in the matrix form are assigned for the corresponding component fields of the $SU(2)_D$ multiplets.
    \label{charges}}
\end{table}

Then, the Lagrangian of our model is given by
\bea
{\cal L}={\cal L}_{\rm DM} +{\cal L}_{\rm VLSM}
\eea
where
the Lagrangian for the dark sector is
\bea
{\cal L}_{\rm DM} &=&-\frac{1}{2} {\rm Tr} \Big(V_{\mu\nu} V^{ \mu\nu}\Big)+i {\bar\Psi} \gamma^\mu D_\mu\Psi +|D_\mu\Phi_D|^2+{\rm Tr}\Big(|D_\mu H'|^2\Big) - V(\Phi_D,H',H) 
\eea
and the Lagrangian for the SM Yukawa couplings including the vector-like leptons is
\bea
{\cal L}_{\rm VLSM}&=&-y_{d} {\bar q}_L d_R  H- y_{u} {\bar q}_L u_R {\tilde H}-y_{l} {\bar l}_L e_R H-y_{\nu} {\bar l}_L \nu_R {\tilde H}-M_R \overline{\nu^c_R}\nu_R   \nonumber \\
&&-M_{E} {\bar \Psi}\Psi-\lambda_{E}  {\bar \Psi}_L\Phi_D  e_{R}-y_{E}  {\bar l}_{L} H' \Psi_{R}  +{\rm h.c.}. \label{leptonL}
\eea
Here, ${\tilde H}=i\sigma^2 H^*$,  $V_{\mu\nu}=\partial_\mu V_\nu-\partial_\nu V_\mu-ig_D[V_\mu,V_\nu]$ with $V_\mu=\frac{1}{2}\tau^i V^i_\mu$ for $SU(2)_D$ gauge bosons, and the covariant derivatives are $D_\mu\Psi=(\partial_\mu - i g_D V_\mu+ig_Y B_\mu)\Psi$, $D_\mu\Phi_D=(\partial_\mu - i g_D V_\mu)\Phi_D$, $D_\mu H=(\partial_\mu - ig  W_\mu-\frac{1}{2} i g_Y B_\mu) H$,  and $D_\mu H'=\partial_\mu H' - ig  W_\mu  H'+ig_D H' V_\mu-\frac{1}{2} i g_Y B_\mu H'$, with $W^\mu=\frac{1}{2}\tau^i W^i_\mu$ and $B_\mu$ being electroweak gauge bosons.

The scalar potential for the singlet scalar $\Phi_D$, the leptophilic Higgs $H'$ and the SM Higgs $H$,  $V(\Phi_D,H,H')$,  is given by
\bea
V(\Phi_D,H, H') &=& \mu^2_1 H^\dagger H + \mu^2_2 {\rm Tr}(H'^\dagger H')-( \mu_3 H^\dagger H' \Phi_D+{\rm h.c.})\nonumber \\
&+& \lambda_1 (H^\dagger H)^2 + \lambda_2 ({\rm Tr}H'^\dagger H')^2+ \lambda_3 (H^\dagger H){\rm Tr}(H'^\dagger H') \nonumber \\
&+& \mu^2_\phi \Phi^\dagger_D\Phi_D + \lambda_\phi (\Phi^\dagger_D\Phi_D)^2+ \lambda_{H\Phi}H^\dagger H\Phi^\dagger_D\Phi_D +  \lambda_{H'\Phi} {\rm Tr}(H'^\dagger H')\Phi^\dagger_D\Phi_D. \label{scalarpot}
\eea
We note that an extra term, $H^\dagger H' {\tilde\Phi}_D$  with $ {\tilde\Phi}_D\equiv i\sigma^2 \Phi^*_D$, is omitted because it violates both $U(1)_G$ and $Z_2$ parity. But, if there is no $Z_2$ parity, we can also introduce $H^\dagger H' {\tilde\Phi}_D$ in our model, but vector dark matter would be unstable.

After the dark $SU(2)_D$ and electroweak symmetries are broken by the VEVs of Higgs fields,
\bea
\langle\Phi_D\rangle= \left( \begin{array}{c}0 \\ v_D\end{array}\right), \,\, \langle H\rangle=\frac{1}{\sqrt{2}} \left( \begin{array}{c}0 \\ v_1\end{array}\right),\,\, \langle H'\rangle=\frac{1}{\sqrt{2}}\left(\begin{array}{cc} 0 & 0 \\ 0 & v_2 \end{array}\right), \label{VEVs}
\eea
the squared mass of the dark gauge bosons becomes 
\bea
m^2_{V^\pm}=m^2_{V^0}=\frac{1}{4} g^2_D (2v^2_D+v^2_2).
\eea
The above tree-level mass relation can be modified due to loop corrections with non-degenerate masses for $SU(2)_D$ doublets, as will be discussed in the next section. Moreover, we get the masses for electroweak gauge bosons as $M_Z=\frac{1}{2} \sqrt{g^2+g^2_Y}\,  v$ and $M_W=\frac{1}{2} g v$, with $v=\sqrt{v^2_1+v^2_2}$, before the mass mixing between $Z$ and $Z'$ gauge bosons is taken into account. The VEV of the SM Higgs doublet $H$ leads to quark masses and mixings, while the VEV of the extra Higgs doublet $H'$ leads to the mixing between the SM leptons and the vector-like lepton as well as the mixing between $Z$ and $Z'$ gauge bosons.

\section{Mass spectra and new interactions}

We discuss the mass spectra for fermions, scalar fields, and show the details of the extra gauge interactions and Yukawa interactions in our model.

\subsection{Fermion masses}

Taking into account the mixing between the vector-like lepton and one lepton, (which we consider as the muon for the later discussion on the muon $g-2$), we first enumerate the mass terms for the lepton sector as
\bea
{\cal L}_{L,{\rm mass}}&=& -M_E {\bar E}E-M_E {\bar E}' E' -m_0 {\bar e}e-( m_R {\bar E}_L e_R+m_L {\bar e}_L E_R+ {\rm h.c.}) \nonumber \\
&=&-M_E {\bar E}' E'- [({\bar e}_L, {\bar E}_L) {\cal M}_L  \left(\begin{array}{c} e_R  \\ E_R \end{array} \right) +{\rm h.c.}]
 \label{leptonmass0}
\eea
where 
\bea
 {\cal M}_L= \left(\begin{array}{cc} m_0 & m_L \\ m_R & M_E \end{array} \right), 
\eea
with $m_0$ being the bare lepton mass given by $m_0=\frac{1}{\sqrt{2}} y_l v_1$, and $m_R, m_L$ being the mixing masses, given by $m_R=\lambda_E v_D$ and $m_L=\frac{1}{\sqrt{2}} y_E v_2$, respectively.

By making a bi-unitary transformation with the rotation matrices for the right-handed leptons and the left-handed leptons, as follows,
\bea
\left(\begin{array}{c} e_L \\  E_L  \end{array}\right)=U_L\left(\begin{array}{c} l_{1L}\\  l_{2L}  \end{array}\right), \quad
\left(\begin{array}{c} e_R \\  E_R \end{array} \right)=U_R\left(\begin{array}{c} l_{1R}\\  l_{2R} \end{array} \right),
\eea
with
\bea
U_L &=&\left(\begin{array}{cc}  \cos\theta_L & \sin\theta_L\\   -\sin\theta_L & \cos\theta_L \end{array} \right), \\
U_R&=& \left(\begin{array}{cc}  \cos\theta_R & \sin\theta_R\\   -\sin\theta_R & \cos\theta_R \end{array} \right),
\eea
we can diagonalize the mass matrix for the vector-like lepton and the lepton as
\bea
U^\dagger_L {\cal M}_L U_R = \left(\begin{array}{cc}  \lambda_- & 0\\   0 & \lambda_+ \end{array} \right).
\eea
Here, we get the mass eigenvalues for leptons \cite{general,kimiko},
\bea
\lambda^2_{-,+}&=& \frac{1}{2} \bigg(m^2_0+ M^2_E +m^2_L +m^{ 2}_R\mp \sqrt{(M^2_E+m^{2}_L-m^2_0-m^2_R )^2+4(m_RM_E+m_L m_0)^2}  \bigg) \nonumber \\
&\equiv& m^2_{l_1,l_2}, \label{masses}
\eea
and the mixing angles as
\bea
\sin(2\theta_R) &=&\frac{2(M_E m_R+m_0 m_L)}{m^2_{l_2}-m^2_{l_1}}, \label{mixR} \\
\sin(2\theta_L) &=&\frac{2(M_E m_L+m_0 m_R)}{m^2_{l_2}-m^2_{l_1}}.   \label{mixL}
\eea

With $\lambda_-=m_{l_1}$ and $\lambda_+=m_{l_2}$ obtained after diagonalization, the lepton mass terms in eq.~(\ref{leptonmass0}) take  the following form,
\bea
{\cal L}_{L,{\rm mass}}= -m_{l_1} {\bar l}_1 l_1 - m_{l_2} {\bar l}_2 l_2 -M_E {\bar E}' E'.
\eea
From eq.~(\ref{masses}), we note that the mass eigenvalues satisfy the following simple relation,
\bea
m_{l_1} m_{l_2} = M_E m_0 -m_R m_L.
\eea
Thus, taking $m_{l_2}\simeq M_E$ for $m_0,m_R, m_L\ll M_E$, we obtain the lepton massses as
\bea
\lambda_-&=&m_{l_1}\approx m_0-\frac{m_R m_L}{M_E}, \label{leptonmass} \\
\lambda_+ &=& m_{l_2}\approx (M^2_E +m^2_L + m^2_R)^{1/2}. \label{hleptonmass}
\eea
Therefore, the seesaw contribution from heavy vector-like leptons \cite{kimiko,seesaw} is naturally small, because it needs
 $m_L\neq 0$ and $m_R\neq 0$, which come from a simultaneous breaking of electroweak symmetry and the $SU(2)_D$ symmetry with the leptophilic Higgs bi-doublet and the dark Higgs doublet.

For $M_E\gg m_0 m_L/m_R, m_0 m_R/m_L$, the mixing angles in eqs.~(\ref{mixR}) and (\ref{mixL}) also get simplified to
\bea
\sin(2\theta_R) &\simeq & \frac{2M_E m_R}{m^2_{l_2}-m^2_{l_1}}\simeq \frac{2m_R}{M_E},  \label{tRapp}\\ 
\sin(2\theta_L) &\simeq & \frac{2M_E m_L}{m^2_{l_2}-m^2_{l_1}}\simeq \frac{2m_L}{M_E}. \label{tRapp}
\eea
Then, together with eq.~({\ref{leptonmass}}), for $m_R, m_L\ll M_E$, we get $ m_{l_1}\sim \frac{m_R m_L}{M_E}\simeq \theta_R \theta_L M_E$, namely, $\theta_R\theta_L\sim m_{l_1}/M_E$. 

In the presence of the $Z-V^0$ mass mixing, as shown in the appendix, and the mass splitting between the vector-like leptons \cite{VQportal,VQportal2} in eq.~(\ref{hleptonmass}), the $SU(2)_D$ gauge bosons have split masses by 
\bea
\Delta m_V\equiv m_{V^0}-m_{V^\pm}\simeq\frac{1}{2}\Delta\rho_H\,\frac{M^2_Z}{m_{V^0}}- \frac{g^2_D m^2_{l_2}}{32\pi^2 m_{V^\pm}} \Big(1-\frac{M^2_{E}}{m^2_{l_2}}\Big)
\eea
where $\Delta\rho_H$ is the tree-level contribution of the $Z-V^0$ mass mixing to the $\rho$ parameter as will be discussed in Section 5, and we took $M_E \gg m_{V^\pm}$ and $m_{V^0}\gg M_Z$.
For instance, from $m^2_{l_2}\simeq M^2_E+m_L^2+m^2_R$ with $m_L\sim \theta_L M_E, m_R\sim \theta_R M_E$, we get the fraction of the mass splitting as
\bea
\frac{\Delta m_V}{m_{V^+}} &\simeq& 2.2\times 10^{-5} \bigg(\frac{\Delta \rho_H}{1.3\times 10^{-3}}\bigg) \bigg(\frac{500\,{\rm GeV}}{m_{V^0}}\bigg)^2 \nonumber \\
&&-1.2\times 10^{-7} \bigg( \frac{g_D}{0.2}\bigg)^2\bigg(\frac{\theta^2_L+\theta^2_R}{2\times 10^{-4}} \bigg)\bigg(\frac{M_E}{1\,{\rm TeV}}\bigg)^2\bigg(\frac{500\,{\rm GeV}}{m_{V^+}}\bigg)^2.
\eea
Therefore, the mass splitting is dominated by the $Z-V^0$ mass mixing, so $\Delta m_V$ tends to be positive, namely, $m_{V^0}> m_{V^+}$.

\subsection{Scalar masses}

First, we consider the minimization of the scalar potential (\ref{scalarpot}) by taking only the neutral components as in eq.~(\ref{VEVs}), but now replacing them with real scalar fields by $v_D\to s$, $v_1\to h_1$ and $v_2\to h_2$. Then, from the scalar potential,
\bea
V(h_1,h_2,s)&=& \frac{1}{2} \mu^2_1 h^2_1 +\frac{1}{2} \mu^2_2 h^2_2 - \mu_3 h_1 h_2 s + \frac{1}{4} \lambda_1 h^4_1  + \frac{1}{4} \lambda_2 h^4_2+ \frac{1}{4} \lambda_3 h^2_1 h^2_2 \nonumber \\
&& + \mu_\phi^2 s^2+\lambda_\phi s^4 +\frac{1}{2}\lambda_{H\Phi} h^2_1 s^2 +\frac{1}{2}\lambda_{H'\Phi} h^2_2 s^2,
\eea 
we obtain the minimization conditions for the VEVs, as follows,
\bea
\frac{\mu_3 v_2 v_D}{v_1}&=&\mu^2_1+\lambda_1 v^2_1 +\frac{1}{2} \lambda_3 v^2_2 + \lambda_{H\Phi} v^2_D, \label{min1} \\
\frac{\mu_3 v_1 v_D}{v_2}&=&\mu^2_2+\lambda_2 v^2_2 +\frac{1}{2} \lambda_3 v^2_1 + \lambda_{H'\Phi} v^2_D, \label{min2} \\
\frac{\mu_3 v_1 v_2}{2v_D}&=&\mu^2_\phi+2\lambda_\phi v^2_D +\frac{1}{2} \lambda_{H\Phi} v^2_1+\frac{1}{2} \lambda_{H'\Phi} v^2_2. \label{min3}
\eea

Next, in order to get the scalar masses, we expand the scalar fields around the VEVs, as follows,
\bea
H=\left(\begin{array}{cc} \phi^+_1 \\ \frac{1}{\sqrt{2}}(v_1+\rho_1+i\eta_1) \end{array} \right), \quad H'=\left(\begin{array}{cc} {\hat\phi}^+_2 & \phi^+_2 \\  {\hat \phi}^0_2 & \frac{1}{\sqrt{2}}(v_2+\rho_2+i\eta_2) \end{array} \right) \label{higgsexpand}
\eea
and
\bea
\Phi_D=\left( \begin{array}{c} \varphi_1 \\  v_D+\frac{1}{\sqrt{2}}(s+i\,a) \end{array}\right). \label{singletexpand}
\eea
Then, we can identify two would-be neutral  Goldstone bosons, $G_Y, G^0_D$, and the CP-odd scalar $A^0$ \cite{seesaw,Bmeson} as 
\bea
G_Y &=& \cos\beta\, \eta_1 +\sin\beta\,\eta_2,  \label{cpodd1} \\
G^0_D&=& \frac{1}{\sqrt{2v^2_D+v^2_2}} \Big(\sqrt{2} v_D\, a-v\sin\beta\,\eta_2\Big),  \label{cpodd2}  \\
A^0&=& N \bigg( \sin\beta\, \eta_1-\cos\beta\,\eta_2 -\frac{v}{\sqrt{2}v_D}\,\sin\beta\cos\beta\, a \bigg)   \label{cpodd3} 
\eea
with
\bea
N= \frac{1}{\sqrt{1+v^2\sin^2\beta\cos^2\beta/(2v^2_D)}},
\eea
and the mass for the CP-odd scalar $A^0$ is given by
\bea
M^2_{A^0} =\frac{\mu_3}{2v_D\sin\beta\cos\beta}\, (v^2\sin^2\beta\cos^2\beta+2v^2_D). \label{CPoddmass}
\eea
Here, we note that $\cos\beta=v_1/v$ and $\sin\beta=v_2/v$ with $v=\sqrt{v^2_1+v^2_2}$.

Moreover, ignoring the mixing between the dark Higgs $s$ and $\rho_{1,2}$, we also obtain the mass eigenstates for CP-even scalars, $h$ and $H$, as
\bea
h &=& \cos\alpha\,\rho_1 +\sin\alpha\, \rho_2, \\
H &=& -\sin\alpha\,\rho_1 + \cos\alpha\, \rho_2
\eea
where $\alpha$ is the mixing angle between CP-even scalars.

We also get the would-be charged Goldstone boson  $G^+$and the charged Higgs $H^+$ by
\bea
G^+ &=& \cos\beta\, \phi^+_1 +\sin\beta\,\phi^+_2, \\
H^+ &=& \sin\beta\,\phi^+_1 -\cos\beta\, \phi^+_2,
\eea
and the mass for the charged Higgs $H^+$ is given by
\bea
M^2_{H^+} = \frac{\mu_3 v_D}{\sin\beta\cos\beta}= \Big(1+\frac{v^2\sin^2\beta\cos^2\beta}{2v^2_D}\Big)^{-1} M^2_{A^0}. \label{Cmass1}
\eea
The extra charged Higgs ${\hat\phi}^+_2$ does not mix with the rest of the scalar fields, with its mass given by
\bea
M^2_{{\hat\phi}^+_2}= \mu_3 v_D \cot\beta=\cos^2\beta\,M^2_{H^+}.  \label{Cmass2}
\eea

Similarly, we also obtain the would-be dark isospin-charged Goldstone boson $G^+_D$ and the dark complex scalar ${\tilde\varphi}$ by
\bea
G^+_D &=& \cos\beta_D\, \varphi_1 +\sin\beta_D\,({\hat\phi}^0_2)^*, \label{nc1}  \\
{\tilde\varphi} &=& \sin\beta_D\,\varphi_1 -\cos\beta_D\, ({\hat\phi}^0_2)^*, \label{nc2}
\eea
with $\cos\beta_D=\sqrt{2}v_D/\sqrt{v^2_2+2v^2_D}$ and $\sin\beta_D=v_2/\sqrt{v^2_2+2v^2_D}$,
and the mass for the dark complex scalar $\tilde\varphi$ is given by
\bea
M^2_{{\tilde\varphi}} = \frac{\mu_3 \cot\beta}{2v_D}\,(v^2_2+2v^2_D)= \cos^2\beta\Big(1+\frac{v^2_2}{2v^2_D}\Big) M^2_{H^+}. \label{Nmass}
\eea

Consequently, we get the mass relation, $M^2_{{\hat\phi}^+_2}\simeq M^2_{H^+}$ for a small $\sin\beta$ from eq.~(\ref{Cmass2}); $M^2_{\tilde\varphi}\simeq M^2_{H^+}$  for $v_2\ll v_D$ from eq.~(\ref{Nmass});  $M^2_{A^0}\gtrsim M^2_{H^+}$ for  $v_D\gtrsim v$ from eq.~(\ref{Cmass1}). 
  
In summary, in the basis of mass eigenstates, we remark that the physical scalar fields are classified into $Z_2$-even and $Z_2$-odd particles:   $h, H, A^0, H^+, s$ are $Z_2$-even, and ${\hat \phi}^+_2, {\tilde\varphi}$ are $Z_2$-odd.

\subsection{Gauge interactions}

 Including the mixing between the muon and the vector-like lepton, we get  the results for the effective interactions of the muon and the vector-like lepton to $SU(2)_D$ gauge bosons and weak gauge bosons as well as the extra gauge interactions for neutral gauge bosons, as follows,
\bea
{\cal L}_{{\rm gauge, eff}} &= &\frac{g_D}{2} V^0_\mu {\bar E}' \gamma^\mu  E' +\bigg(\frac{g_D}{\sqrt{2}} V^+_\mu {\bar E}'_L \gamma^\mu  (-s_L \mu_L + c_L E_L) +{\rm h.c.} \bigg)
\nonumber \\
&&+\bigg(\frac{g_D}{\sqrt{2}} V^+_\mu  {\bar E}'_R \gamma^\mu  (-s_R \mu_R+ c_R E_R) +{\rm h.c.} \bigg)+\frac{g}{2c_W}(v_l-a_l) Z_\mu  {\bar E}' \gamma^\mu E'\nonumber \\
&&-\frac{1}{2} g_D V^0_\mu \Big(c^2_R {\bar E} \gamma^\mu P_R  E+ s^2_R\,  {\bar \mu} \gamma^\mu P_R \mu -s_Rc_R ({\bar E}\gamma^\mu P_R \mu + {\bar \mu}\gamma^\mu P_R E) \nonumber \\
&&\quad+c^2_L {\bar E} \gamma^\mu  P_L E+s^2_L {\bar \mu} \gamma^\mu P_L \mu -s_Lc_L ( {\bar E}\gamma^\mu P_L \mu+ {\bar \mu}\gamma^\mu P_L E)   
  \Big) \nonumber \\
&&+\frac{g}{2c_W}\, Z_\mu (v_l+a_l) \Big( c^2_L{\bar \mu} \gamma^\mu P_L \mu + s_Lc_L( {\bar E}\gamma^\mu P_L \mu+ {\bar \mu}\gamma^\mu P_L E)+s^2_L{\bar E} \gamma^\mu P_L E    \Big) \nonumber \\
&&+\frac{g}{2c_W}\, Z_\mu (v_l-a_l) \Big( {\bar \mu}\gamma^\mu P_R \mu + s^2_L {\bar \mu}\gamma^\mu P_L \mu - s_Lc_L( {\bar E}\gamma^\mu P_L \mu+ {\bar \mu}\gamma^\mu P_L E)\nonumber \\
&&\quad + {\bar E} \gamma^\mu P_R  E+c^2_L {\bar E} \gamma^\mu P_L E \Big)  \nonumber \\
&&+\frac{g}{\sqrt{2}}\, W^-_\mu\Big( c_L{\bar \mu}\gamma^\mu P_L \nu_\mu+ s_L {\bar E} \gamma^\mu P_L \nu_\mu \Big) +{\rm h.c.} +{\cal L}_{Z-{\rm mix}}  \label{gauge}
\eea
where the lepton mixing angles are $s_R=\sin\theta_R, c_R=\cos\theta_R$, $s_L=\sin\theta_L, c_L=\cos\theta_L$, the Weinberg angle is denoted by $s_W=\sin\theta_W, c_W=\cos\theta_W$, $v_l=\frac{1}{2}(-1+4s^2_W)$,  $a_l=-\frac{1}{2}$, and  we used the same notations for the mass eigenstates for leptons as those for the interaction eigenstates, $\mu, E, E'$, for simplicity. We note that $V^\pm$ and $E'$ carry $Z_2$-odd parities, so the lighter particle would be stable. Then, for $M_{E}\gtrsim m_{V^\pm}$, the dark isospin charged gauge bosons, $V^\pm$, can be candidates for stable dark matter.  

We note that ${\cal L}_{Z-{\rm mix}}$ in eq.~(\ref{gauge}) is the Lagrangian for the extra gauge interactions for the SM fermions and the vector-like lepton, due to the mass mixing between $Z$  and $V^0$ gauge bosons.
For small lepton mixing angles, it is given by 
\bea
{\cal L}_{Z-{\rm mix}} &=&\frac{1}{2} g_D ((c_\zeta-1)V^0_\mu -s_\zeta Z_\mu)({\bar E}'\gamma^\mu E'- {\bar E}\gamma^\mu E )   \nonumber \\
&&+\frac{e}{2c_W s_W}\sum_{f={\rm SM}, E,E'} (s_\zeta V^0_\mu+ (c_\zeta-1) Z_\mu) \Big[{\bar f}\gamma^\mu (\tau^3-2 s^2_W Q_f) f\Big]
\eea
where $c_\zeta=\cos\zeta, s_\zeta=\sin\zeta$, $\zeta$ is the mixing angle between $Z$ and $V^0$ gauge bosons \cite{Bmeson,seesaw,kimiko}, given in eq.~(\ref{Zmix}), and the mass eigenvalues for  $Z_1$($Z$-like) and $Z_2$($V^0$-like) gauge bosons, $M_{Z_{1,2}}$, are given in eq.~(\ref{Zmasses}). Here, we denoted the mass eigenstates for the massive neutral gauge bosons by those for the interaction eigenstates, $Z, V^0$, for simplicity. 
Unlike in the case with a $U(1)'$ gauge symmetry \cite{Bmeson,seesaw,kimiko}, there is no gauge kinetic mixing allowed for the $SU(2)_D$ gauge symmetry in our case. The mass mixing between $Z$ and $V^0$ gauge bosons leads to interesting signatures such as dilepton searches at the LHC and the electroweak precision observables \cite{kimiko}, and the correction to the $W$ boson mass, as will be discussed in the later section.

We remark that the relevant effective vector and axial-vector couplings to $V^0$ and $Z$ containing the lepton and the vector-like lepton  are given by
\bea
{\cal L}_{{\rm gauge, eff}}&\supset& -g_D V^0_\mu {\bar \mu}\gamma^\mu (v'_\mu+a'_\mu \gamma^5) \mu + \Big(g_D V^0_\mu {\bar \mu} \gamma^\mu (c_V+c_A \gamma^5)E +{\rm h.c.}\Big) \nonumber \\
&&-\frac{g_D}{\sqrt{2}} V^-_\mu {\bar \mu} \gamma^\mu({\hat c}_V+{\hat c}_A\gamma^5 )E' +{\rm h.c.} \nonumber \\ 
&&+\frac{e}{2c_W s_W} c_\zeta Z_\mu\,  {\bar \mu}\gamma^\mu (v_l-a_l \gamma^5) \mu \nonumber \\
&&+\frac{g}{2c_W}\,(a_l s_L c_L)Z_\mu\, {\bar \mu} \gamma^\mu(1-\gamma^5)E  +{\rm h.c.}  \label{gaugesum}
\eea
where
\bea
v'_\mu&=& \frac{1}{4} (\sin^2\theta_R+\sin^2\theta_L)-\frac{e s_\zeta v_l}{2c_W s_W g_D},  \label{ve} \\
a'_\mu&=& \frac{1}{4}(\sin^2\theta_R-\sin^2\theta_L) +\frac{e s_\zeta a_l}{2c_W s_W g_D}, \label{ae} \\
c_V&=& \frac{1}{8} (\sin2\theta_R+\sin2\theta_L), \label{cv}  \\
c_A &=& \frac{1}{8} (\sin2\theta_R-\sin2\theta_L), \label{ca} \\
{\hat c}_V &=&\frac{1}{2}(\sin\theta_L+\sin\theta_R) , \label{hve} \\
{\hat c}_A &=&-\frac{1}{2}(\sin\theta_L-\sin\theta_R). \label{hae} 
\eea
The $V^0, V^\pm$ interactions for transitions between the lepton and the vector-like lepton can be used to explain the muon $g-2$ anomaly by one-loop corrections, as will be discussed later.

We also get the  $SU(2)_D$ self-interactions and the gauge interactions of the SM Higgs $h$ and the singlet scalar $s$ as
\bea
{\cal L}_{\rm gauge} &=&{\cal L}_{\rm self}+{\cal L}_{\rm scalars}
\eea
where
\bea
{\cal L}_{\rm self}&=& -i g_D\bigg[
\left(\partial^\mu V^{-\nu} -\partial^\nu V^{-\mu}\right)
 V^+_\mu V^0_\nu -\left(\partial^\mu V^{+\nu} -\partial^\nu V^{+\mu}\right)
 V^-_\mu V^0_\nu \nonumber \\
 &&\quad + V^-_\mu V^+_\nu\Big( \partial^\mu V^{0\nu} - \partial^\nu V^{0\mu}\Big)
\bigg]-{g^2_D\over 2}\left[
\left(V^-_\mu V^{+\mu}\right)^2 - V^-_\mu V^{-\mu}
V^+_\nu V^{+\nu} \right]   \nonumber \\
&& - g^2_D
V^-_\mu V^{+\mu}  V^0_{\nu} V^{0\nu} +g^2_D V^-_\mu  V^+_\nu  V^{0\mu} V^{0\nu},
\label{dgauge}
\eea
and 
\bea
{\cal L}_{\rm scalars}&=&\frac{1}{8} g^2_D \, s(s+2\sqrt{2} v_D) (V^0_\mu V^{0\mu} +2 V^+_\mu V^{-\mu} ) \nonumber \\
&&+\frac{1}{8}\sin^2\beta\,  h(2v+h) \bigg( g^2_D  (V^0_\mu V^{0\mu} + 2 V^-_\mu V^{+\mu}) -\frac{2g g_D}{c_W} \, V^0_\mu Z^\mu \bigg) \nonumber \\
&&+\frac{1}{8} g^2\, h(2v+h)\bigg(\frac{1}{c^2_W}\, Z_\mu Z^\mu +2 W_\mu W^\mu\bigg). \label{scalars}
\eea
Here, we took the alignment limit for the SM Higgs and decoupled the extra scalars.
For the mass mixing between neutral gauge bosons, we only have to replace $V^0_\mu$ by $c_\zeta V^0_\mu -s_\zeta Z_\mu$ in eqs.~(\ref{dgauge}) and ~(\ref{scalars}).

\subsection{Yukawa interactions}
 
We present the Yukawa interactions for neutral and charged scalar fields  in our model and identify the relevant interactions for the muon $g-2$.

After expanding the scalar fields around the VEVs as in eqs.~(\ref{higgsexpand}) and (\ref{singletexpand}), we get the Yukawa couplings for quarks and leptons with the SM Higgs, and extra neutral scalar fields and the charged scalar,  as follows \cite{seesaw}, 
\bea
{\cal L}_Y={\cal L}_q+{\cal L}_l
\eea
where
\bea
{\cal L}_q&=& - \sum_{f=u,d} \frac{m_f}{v} \,\bigg(\xi^f_h {\bar f} f  h+\xi^f_H {\bar f} f H -i\xi^f_A {\bar f} \gamma_5 f A^0 \bigg) \nonumber \\
&&+\frac{\sqrt{2}}{N\, v}\, V_{ud}\, {\bar u} \Big(m_u\, \xi^u_A P_L +m_d\, \xi^d_A P_R \Big) d\, H^++{\rm h.c.}, \label{effYq}
\eea
with
\bea
\xi^u_h &=& \xi^d_h = \frac{\cos\alpha}{\cos\beta}, \label{hquark} \\
\xi^u_H &=& \xi^d_H = -\frac{\sin\alpha}{\cos\beta}, \\
\xi^u_A &=& -\xi^d_A =N \tan\beta,
\eea
and
\bea
{\cal L}_l &=&-\frac{m_0}{v} \, (c_L{\bar e}+s_L {\bar E}) \Big( \xi^d_h\, h+\xi^d_H\, H-i \xi^d_A \, A\Big) P_R (c_R e +s_R E) +{\rm h.c.} \nonumber \\
&&-\frac{m_L}{v}\, (c_L{\bar e}+s_L {\bar E}) \Big( \xi^l_h \, h + \xi^l_H  \, H-i \xi^{l_L}_A \,  A\Big) P_R(c_R E- s_R e) +{\rm h.c.} \nonumber \\
&& +\frac{m_L}{v}\,  i \,\xi^{l_R}_A\, (c_R{\bar e}+s_R{\bar E}) P_L (c_L E-s_L e)  A^0+{\rm h.c.} \nonumber \\
&&-\frac{m_R}{\sqrt{2} v_D}\, (c_R {\bar e}+s_R {\bar E}) P_L (c_LE-s_L e) s+{\rm h.c.} \nonumber \\
&&+\frac{\sqrt{2}m_0}{N v} \xi^d_A  {\bar\nu}P_R (c_R {\bar e}+s_R {\bar E}) H^+ +{\rm h.c.}  \nonumber \\
&& +\frac{\sqrt{2}m_L}{N \,v} \, {\bar \nu}\, \xi^{l_L}_A P_R (c_R E-s_R e) H^+ + \frac{\sqrt{2}m_L}{N \,v} \, {\bar \nu}\, \xi^{l_L}_A P_R  E' {\hat\phi}^+_2 +{\rm h.c.} \nonumber \\
&& - \frac{m_R}{v_D}\,\sin\beta_D (c_R{\bar e} +s_R {\bar E}) P_L E' {\tilde\varphi} +\frac{\sqrt{2} m_L}{v}\,\frac{\cos\beta_D}{\sin\beta}\,(c_L {\bar e}+ s_L {\bar E})P_R E' {\tilde\varphi}  +{\rm h.c.},
\label{effYukawa}
\eea
with
\bea
\xi^l_h &=& \frac{\sin\alpha}{\sin\beta}, \label{hlepton} \\
\xi^l_H &=&  \frac{\cos\alpha}{\sin\beta}, \\
\xi^{l_L}_A &=& N \cot\beta, \\
\xi^{l_R}_A &=& -\frac{1}{2}N  \Big(\frac{m_R}{m_L}\Big) \Big( \frac{v}{v_D}\Big)^2 \sin\beta\cos\beta.
\eea
Here, we used the same notations for the mass eigenstates for leptons as for interaction eigenstates.
 
In the alignment limit for the Higgs sector with $\alpha=\beta$, we get $\xi^u_h=\xi^d_h=1$ from eq.~(\ref{hquark}), which leads to the Higgs Yukawa couplings for quarks as in the SM. Similarly, from the first two lines of eq.~(\ref{effYukawa}) with $\alpha=\beta$,  we also obtain the Higgs Yukawa couplings for charged leptons by
\bea
{\cal L}_{h-l-l} &=& -\frac{c_L}{v} \bigg(\frac{m_0\cos\alpha}{\cos\beta}-\frac{m_L\sin\alpha}{\sin\beta}\, s_R \bigg) h\, {\bar e} e \nonumber \\
&\simeq & -\frac{1}{v} \bigg( m_0 -\frac{m_L m_R}{M_E}\bigg)h\, {\bar e} e\simeq - \frac{m_l}{v} \, h\, {\bar e} e
\eea
where we used $c_L\simeq 1$, $s_R\simeq m_R/M_E$ from eq.~(\ref{tRapp}), and the seesaw lepton mass, $m_l\simeq m_0-m_L m_R/M_E$, from eq.~(\ref{leptonmass}).
In this case, we can recover the Higgs Yukawa couplings for quarks and leptons as in the SM. 
  
As a result, for the muon $g-2$, we list  the effective Yukawa interactions  for neutral real scalar fields,  $h_i=h, H, s, A^0 \, (i=1,2,3,4)$, a  neutral complex scalar field, $h_5= {\tilde\varphi}$, and charged scalar $H^-$, containing the lepton and/or the vector-like lepton, as follows,
 \bea
 {\cal L}_l &\supset&  - \Big(\sum_{i=1}^4{\bar \mu}  (v^E_i- ia^E_i \gamma^5 )E  h_i +{\bar \mu}  (v^E_5- ia^E_5 \gamma^5 )E'  h_5 +{\rm h.c.}\Big) \nonumber \\
&& -\sum_{i=1}^4 {\bar \mu}  (v^\mu_i- ia^\mu_i \gamma^5 ) \mu  h_i  - \Big(v_{H^-} {\bar \mu}  (1-  \gamma^5 )\nu_\mu H^- +{\rm h.c.}\Big)
\eea
where
\bea
v^E_1&=& \frac{m_L}{2v}\,(c_L c_R-s_L s_R)\xi^\mu_h, \\
-ia^E_1&=&  \frac{m_L}{2v}\,(c_L c_R+s_L s_R)\xi^\mu_h, \\
v^E_2 &=&  \frac{m_L}{2v}\,(c_L c_R-s_L s_R)\xi^\mu_H, \\ 
 -ia^E_2&=& \frac{m_L}{2v}\,(c_L c_R+s_L s_R)\xi^\mu_H , \\
v^E_3 &=&  \frac{m_R}{2\sqrt{2} v_D}(c_L c_R-s_L s_R), \\
-i a^E_3&=& -\frac{m_R}{2\sqrt{2} v_D}(c_L c_R+s_L s_R), \\
v^E_4 &=& - \frac{i m_L}{2v}\,(c_L c_R+s_L s_R)(\xi^{\mu_L}_A+\xi^{\mu_R}_A),\\
a^E_4 &=&\frac{m_L}{2v}\,(c_L c_R-s_L s_R)(\xi^{\mu_L}_A-\xi^{\mu_R}_A),  \\
v^E_5 &=& \frac{m_R}{2v_D} \sin\beta_D \, c_R-\frac{\sqrt{2} m_L}{2v} \,\frac{\cos\beta_D}{\sin\beta}\,c_L, \\
-ia^E_5 &=& - \frac{m_R}{2v_D} \sin\beta_D \, c_R-\frac{\sqrt{2} m_L}{2v} \,\frac{\cos\beta_D}{\sin\beta}\,c_L, 
\eea
and
\bea
v^\mu_1 &=&\frac{m_0}{v} \,c_L c_R\, \xi^d_h-\frac{m_L}{v}\,c_L s_R\, \xi^\mu_h, \\
v^\mu_2 &=&\frac{m_0}{v}\, c_L c_R\, \xi^d_H -\frac{m_L}{v}\,c_L s_R\, \xi^\mu_H, \\
v^\mu_3 &=&-\frac{m_R}{\sqrt{2}v_D}\,c_R s_L,  \\ 
a^\mu_4 &=&\frac{m_0}{v}\, c_L c_R\, \xi^d_A -\frac{m_L}{v} (c_L s_R\, \xi^{\mu_L}_A -c_R s_L \, \xi^{\mu_R}_A), \\
v_{H^-} &=&-\frac{\sqrt{2}m_0}{2N v}\, c_R\, \xi^d_A+ \frac{\sqrt{2}m_L}{2 N v}\,s_R\,\xi^{\mu_L}_A,
\eea
and $v^\mu_4=a^\mu_{1,2,3}=0$.

\section{Muon $g-2$}

From the combined average with Brookhaven E821 \cite{Muong-2:2006rrc}, the difference from the SM prediction \cite{Aoyama:2020ynm}  becomes
\bea
\Delta a_\mu = a^{\rm exp}_\mu -a^{\rm SM}_\mu =251(59)\times 10^{-11}, \label{amu-recent}
\eea
showing a $4.2\sigma$  discrepancy from the SM \cite{Muong-2:2021ojo}.
We remark that the recent lattice-QCD calculation with reduced errors shows a smaller deviation that the value quoted above \cite{bmw}. In this case, however, it was noticed that the $e^+e^-$ hadronic cross section below about $1\,{\rm GeV}$ would be in tension with the experimental data \cite{passera,epemdata} and the global electroweak fit \cite{passera1,passera,ewfit}.
In the following discussion, we take the results shown in eq.~(\ref{amu-recent}) as face value and discuss the implications of our model for the corrections to the muon $g-2$. 

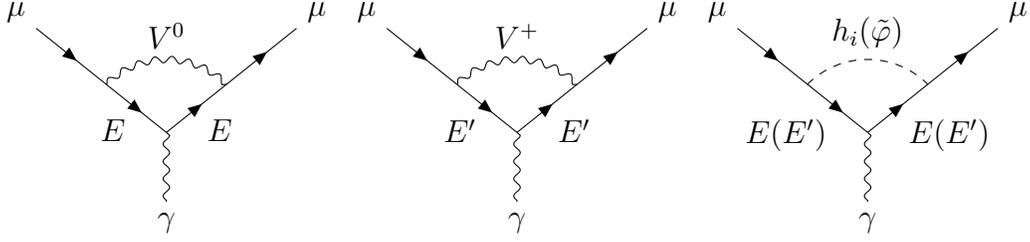
\begin{figure}[!t]
\begin{center}
\begin{tikzpicture}[baseline=($0.5*(ein)+0.5*(pho)$)]
		\begin{feynman}[inline=($0.5*(ein)+0.5*(pho)$),medium]
			\vertex (ein) at (-2.0, 1.2) {\( \mu \)};
			\vertex (pho) at (0, -1.6) {\( \gamma \)};
			\vertex (eout) at (2.0, 1.2) {\( \mu \)};
			\vertex (int1) at (-0.8, 0.24);
			\vertex (int2) at (0, -0.4);
			\vertex (int3) at (0.8, 0.24);
		
			\diagram* {
				(ein) -- 
				[fermion] (int1) -- 
				[fermion, edge label' = \( E \)] (int2) -- 
				[fermion, edge label' = \( E \)] (int3) -- 
				[fermion] (eout) ,
				(pho) -- [photon] (int2),
				(int1) -- [photon, quarter left, edge label = \( V^0 \)] (int3),
			};
		\end{feynman}
	\end{tikzpicture}
\begin{tikzpicture}[baseline=($0.5*(ein)+0.5*(pho)$)]
		\begin{feynman}[inline=($0.5*(ein)+0.5*(pho)$),medium]
			\vertex (ein) at (-2.0, 1.2) {\( \mu \)};
			\vertex (pho) at (0, -1.6) {\( \gamma \)};
			\vertex (eout) at (2.0, 1.2) {\( \mu \)};
			\vertex (int1) at (-0.8, 0.24);
			\vertex (int2) at (0, -0.4);
			\vertex (int3) at (0.8, 0.24);
		
			\diagram* {
				(ein) -- 
				[fermion] (int1) -- 
				[fermion, edge label' = \( E' \)] (int2) -- 
				[fermion, edge label' = \( E' \)] (int3) -- 
				[fermion] (eout) ,
				(pho) -- [photon] (int2),
				(int1) -- [photon, quarter left, edge label = \( V^+ \)] (int3),
			};
		\end{feynman}
	\end{tikzpicture}
\begin{tikzpicture}[baseline=($0.5*(ein)+0.5*(pho)$)]
		\begin{feynman}[inline=($0.5*(ein)+0.5*(pho)$),medium]
			\vertex (ein) at (-2.0, 1.2) {\( \mu \)};
			\vertex (pho) at (0, -1.6) {\( \gamma \)};
			\vertex (eout) at (2.0, 1.2) {\( \mu \)};
			\vertex (int1) at (-0.8, 0.24);
			\vertex (int2) at (0, -0.4);
			\vertex (int3) at (0.8, 0.24);
		
			\diagram* {
				(ein) -- 
				[fermion] (int1) -- 
				[fermion, edge label' = \( E (E') \)] (int2) -- 
				[fermion, edge label' = \( E (E') \)] (int3) -- 
				[fermion] (eout) ,
				(pho) -- [photon] (int2),
				(int1) -- [scalar, quarter left, edge label = \( h_i (\tilde{\varphi}) \)] (int3),
			};
		\end{feynman}
	\end{tikzpicture}	
\end{center}
\caption{Feynman diagrams for one-loop corrections to the muon $g-2$. }
\label{g2Feyn}
\end{figure}

In our model,  the vector-like leptons and the $SU(2)_D$  gauge bosons as well as extra scalars contribute to the muon $g-2$ at one loop, as follows,
\bea
\Delta a_\mu= \Delta a^{V,E}_\mu+\Delta a^{Z,E}_\mu+ \Delta a^{V,\mu}_\mu  +\Delta a^{h,E}_\mu +\Delta a^{h,\mu}_\mu +\Delta a^{H^-}_\mu.
\eea
As shown in Fig.~\ref{g2Feyn}, the transitions between the SM lepton and the vector-like lepton via $V^0, V^\pm$,  $s$ or ${\tilde\varphi}$, give rise to the dominant contributions to the muon $g-2$ by $ \Delta a^{V,E}_\mu$ and $\Delta a^{h,E}_\mu$.
As compared to the $U(1)'$ case \cite{seesaw,kimiko},  there are new contributions coming from the extra gauge bosons ($V^\pm$) of $SU(2)_D$, the vector-like lepton $E'$ in the $SU(2)_D$ doublet and the complex scalar ${\tilde\varphi}$, which is the combination of the neutral scalars in the Higgs bi-doublet and in the $SU(2)_D$ Higgs doublet.

In this section, we present the extra contributions to the muon $g-2$ in our model and check the consistent parameter space, applicable for the analysis on the $W$ boson mass and dark matter constraints in the later sections.

\subsection{Vector contributions}

We consider the one-loop corrections to the muon $g-2$ in our model in detail. 
First, we list $ \Delta a^{V,E}_\mu $ as the contributions from the $SU(2)_D$ gauge bosons and the vector-like leptons running together in loops  \cite{kimiko,seesaw}, given by
\bea
\Delta a^{V,E}_{\mu} =\Delta a^{V^0,E}_\mu+\Delta a^{V^\pm,E'}_\mu
\eea
where
\bea
\Delta a^{V^0,E}_\mu&=& \frac{ g^2_D m_\mu}{4\pi^2 m_{V^0}}r_E \bigg[ c^2_{V} \bigg\{ \frac{m_{\mu}}{m_{V^0}} \Big( \frac{5}{6} -\frac{5}{2}r_E+r^2_E +(r^3_E-3r^2_E+2r_E) \ln \frac{r_E-1}{r_E}\Big) \nonumber \\
&&  +\frac{M_E}{m_{V^0}}  \Big(2r_E-1 +2(r^2_E-r_E)  \ln \frac{r_E-1}{r_E}\Big)  \nonumber \\
&&+\frac{m_{\mu} M^2_E }{2m^3_{V^0}} \Big(\frac{5}{6} +\frac{3}{2} r_E +r^2_E +(r^2_E+r^3_E) \ln \frac{r_E-1}{r_E}  \Big) \nonumber \\
&& -\frac{M^3_E}{2m^3_{V^0}} \Big( \frac{1}{2} +r_E+r^2_E\ln \frac{r_E-1}{r_E} \Big) \bigg\} +c^2_{A} \{M_E\to -M_E \} \bigg] \label{amu}
\eea
with 
\bea
r_E=\bigg(1-\frac{M^2_E}{m^2_{V^0}} \bigg)^{-1},
\eea
and 
\bea
\Delta a^{V^\pm,E'}_\mu =\frac{1}{2} \Delta a^{V^0,E}_l(c_V\to {\hat c}_V, c_A\to {\hat c}_A, M_E\to M_{E'}, m_{V^0}\to m_{V^\pm}).
\eea
We note that $c_V, c_A$ and ${\hat c}_V, {\hat c}_A$ are given in eqs.~(\ref{cv}), (\ref{ca}), (\ref{hve}) and (\ref{hae}).
Here, we took $m_{V^0}\simeq m_{V^\pm}$, and $m_{l_2}\simeq M_E\simeq M_{E'}$, $m_{l_1}= m_\mu$ for $m_L, m_R\ll M_E$.
Then, the above results can be approximated to
\bea
\Delta a^{V,E}_{\mu} \simeq  \left\{\begin{array}{c} \frac{g^2_D M_{E} m_{\mu}}{16 \pi^2 m_{V^0}^2}\, (c_{V}^2-c_{A}^2)+ \frac{g^2_D M_{E} m_{\mu}}{32 \pi^2 m_{V^0}^2}\, ({\hat c}_{V}^2-{\hat c}_{A}^2), \qquad M_{E}\gg m_{V^0}, \vspace{0.3cm} \\ \frac{g^2_D M_{E} m_{\mu}}{4 \pi^2 m_{V^0}^2}\, (c_{V}^2-c_{A}^2)+\frac{g^2_D M_{E} m_{\mu}}{8 \pi^2 m_{V^0}^2}\, ({\hat c}_{V}^2-{\hat c}_{A}^2), \quad m_{\mu}\ll M_{E}\ll m_{V^0}. \end{array}  \right. \label{zp1}
\eea
As a result, from eqs.~(\ref{cv}) and (\ref{ca}), we find that $c_V>c_A$, which leads to $\Delta a^{V^0,E}_\mu>0$.
Since $c_{V}\simeq \frac{1}{4}(\theta_{R}+\theta_{L})$ and $c_{A}\simeq \frac{1}{4}(\theta_{R}-\theta_{L})$, with $\theta_{R}\simeq \frac{m_{\mu}}{m_{L}}$ and $\theta_{L}\simeq  \frac{m_{L}}{M_{E}}$, we find that $c_{V}^2-c_{A}^2\simeq \frac{1}{4}\theta_{L} \theta_{R}\simeq \frac{m_{\mu}}{4M_{E}}$.   Similarly, from eqs.~(\ref{hve}) and (\ref{hae}), we get ${\hat c}_V\simeq \frac{1}{2}(\theta_L+\theta_R)$ and ${\hat c}_A\simeq -\frac{1}{2}(\theta_L-\theta_R)$, leading to ${\hat c}_{V}^2-{\hat c}_{A}^2\simeq \theta_R\theta_L\simeq \frac{m_\mu}{M_E}$. 
Therefore, we find that the large  chirality-flipping effect from $M_{E}$ is cancelled by the small mixing angles such that $\Delta a^{V,E}_{l}\simeq \frac{3g^2_D}{64\pi^2}  \frac{m^2_{\mu}}{m^2_{V^0}}$ for $M_E\gg m_{V^0}$, and it shows a non-decoupling phenomenon, almost independent of $M_{E}$.

Similarly, $\Delta a^{Z,E}_\mu$ is the contribution coming from the $Z$-boson and the vector-like lepton, approximated for $M_{E}\gg m_Z$ to
\bea
\Delta a^{Z,E}_{\mu}\simeq  -\frac{5g^2 a^2_l s^2_L c^2_L m^2_{\mu}}{ 96 \pi^2 c^2_W m_{Z}^2}\,\Big(1+\frac{6 m^2_{Z}}{5 M^2_{E}} \Big),
\eea
which is negative but subdominant for the small mixing angles for the vector-like lepton.

Likewise, $ \Delta a^{V^0,\mu}_\mu $ is the $(g-2)_\mu$ contribution from $V^0$ and muon running together in loops, approximated for  $m_{Z'}\gg m_\mu$ to
\bea
\Delta a^{V^0,\mu}_{\mu} &=& \frac{g^2_D m^2_\mu}{4\pi^2} \int^1_0 dx \,  \frac{\big[v^{\prime 2}_\mu x^2(1-x)-a^{\prime 2}_\mu \big(x(1-x)(4-x) +\frac{2m^2_\mu x^3}{m^2_{V^0}}\big)\big]}{m^2_\mu x^2+m^2_{V^0}(1-x)} \nonumber \\
&\simeq& \frac{g^2_D m^2_\mu}{12\pi^2 m^2_{V^0}}\, ( v^{\prime 2}_{\mu} -5 a^{\prime 2}_{\mu})
\eea
where we assumed  $m_{V^0}\gg m_\mu$ in the second line, and $v'_\mu, a'_\mu$ are given in eqs.~(\ref{ve}) and (\ref{ae}), respectively.  Thus, we find that $ \Delta a^{V^0,\mu}_{\mu} $ is sub-dominant for the small mass mixing for the neutral gauge bosons (i.e. $|\sin\zeta|\ll 1$) and the small mixing angles for the vector-like lepton (i.e. $\theta_{R,L}\ll 1$).

\begin{figure}[t]
\centering
\includegraphics[width=0.43\textwidth,clip]{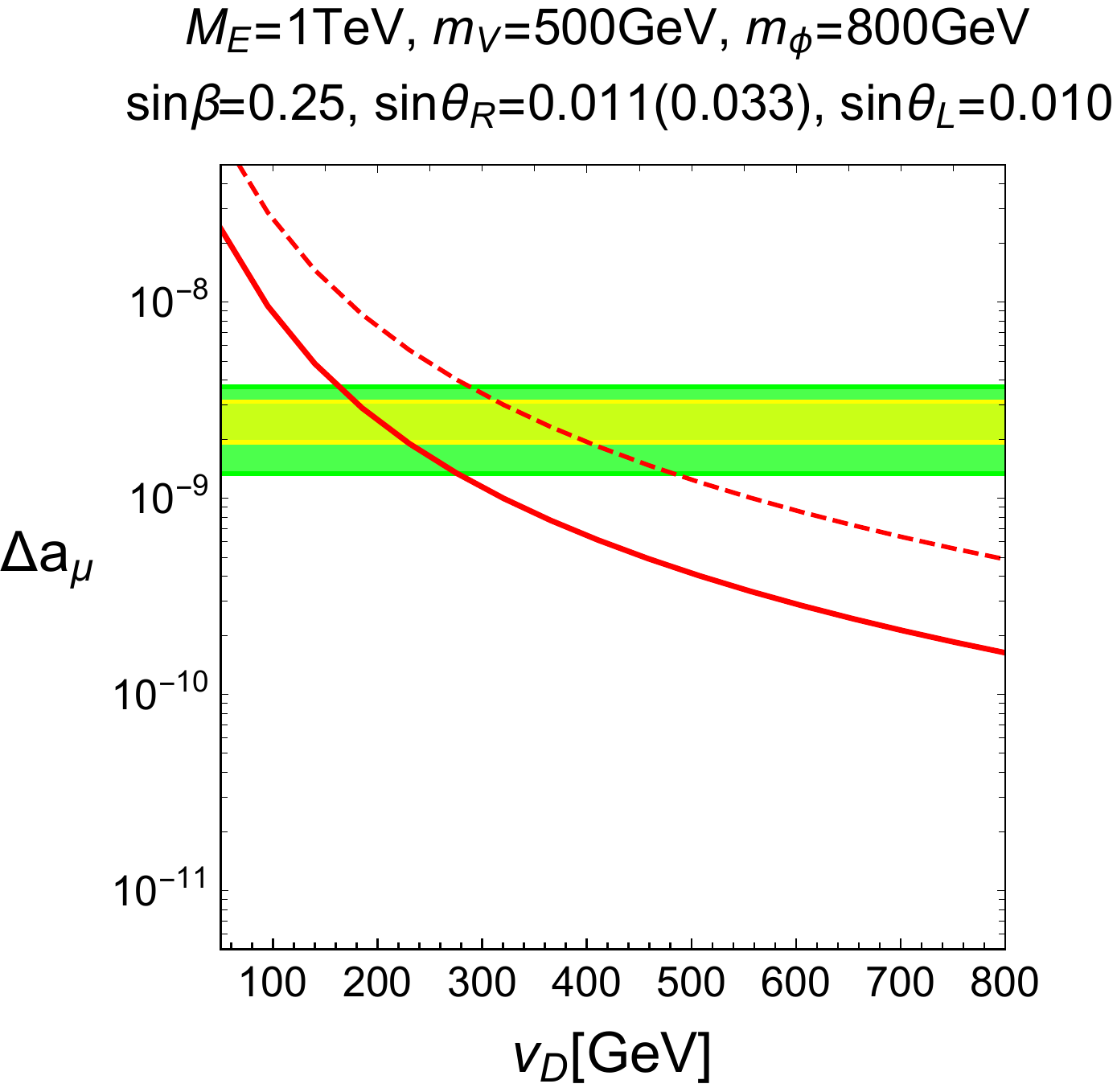}\,\,\,\,\,\,
\includegraphics[width=0.43\textwidth,clip]{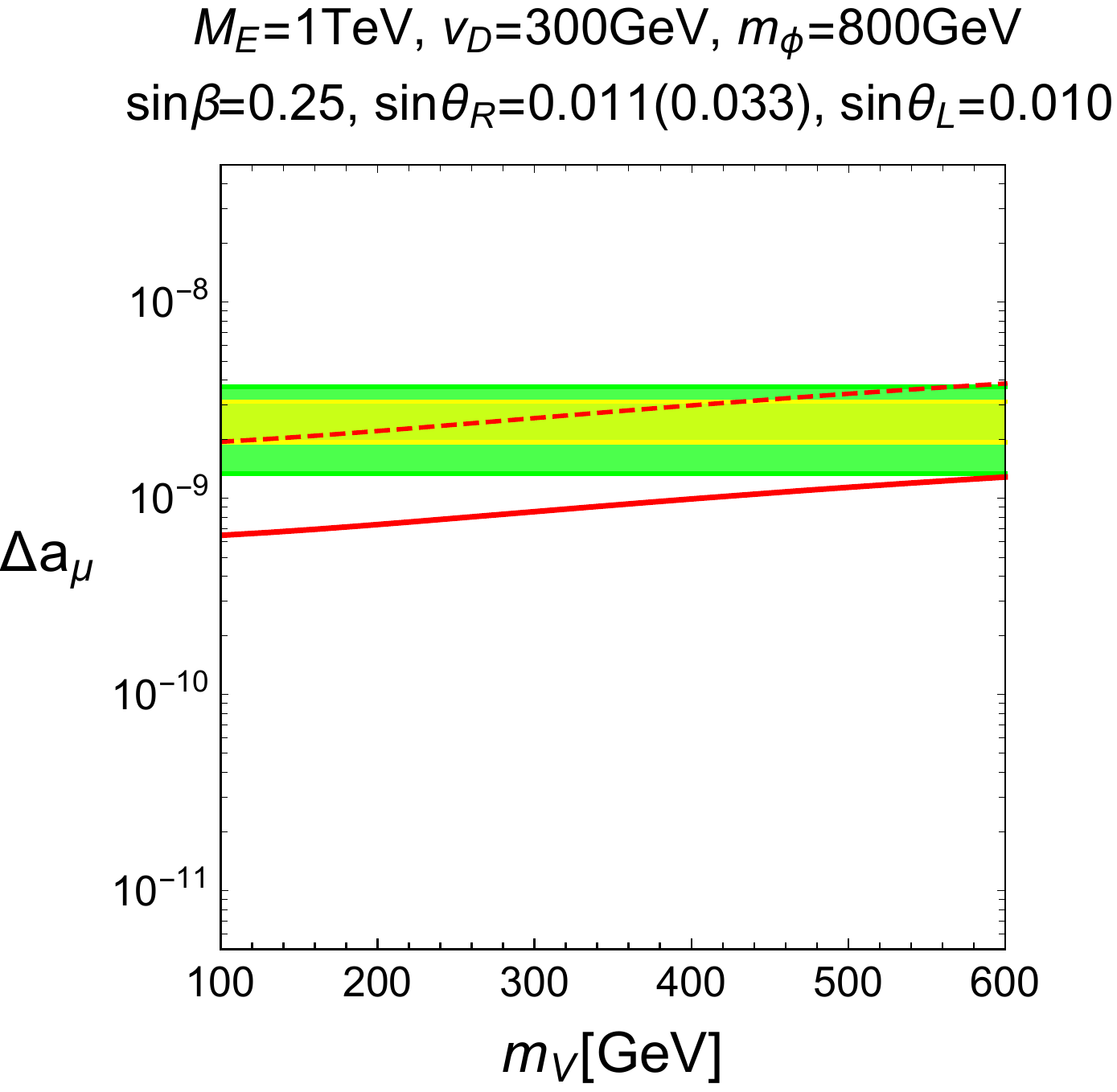}
\caption{(Left) $\Delta a_{\mu}$  as a function of $v_D$ in comparison to the $1\sigma$($2\sigma$) bands for the deviation of the muon $g-2$ \cite{Muong-2:2021ojo} in yellow(green).  (Right) $\Delta a_{\mu}$  as a function of $m_{V^+}$.  For both plots, we chose $M_E=1\,{\rm TeV}$, $m_{\tilde\phi}=800\,{\rm GeV}$ and $\sin\beta=0.25$, and the full one-loop contributions to the muon $g-2$ for $\sin\theta_R=0.011 (0.033)$ and $\sin\theta_L=0.010$ are shown in red solid(dashed) lines. We also chose $m_V=500\,{\rm GeV}$ on left and $v_D=300\,{\rm GeV}$ on right. 
}
\label{fig:g-2}
\end{figure}

Similarly, we also remark that the $Z$-boson contributions to the muon $g-2$ can be corrected due to the modified neutral current interaction for a nonzero mixing between $Z$ and $V^0$ gauge bosons, as shown in the third line in eq.~(\ref{gaugesum}). Then, we find that the SM contribution with $Z$ and muon running together in loops deviates by $ \Delta a^{Z,\mu}_\mu $, which is given by
\bea
\Delta a^{Z,\mu}_\mu&=& \frac{e^2 m^2_\mu}{16\pi^2 c^2_W s^2_W} (c^2_\zeta-1) \int^1_0 dx \,  \frac{\big[v^2_\mu x^2(1-x)-a^{2}_\mu \big(x(1-x)(4-x) +\frac{2m^2_\mu x^3}{M^2_Z}\big)\big]}{m^2_\mu x^2+M^2_Z(1-x)} \nonumber \\
&\simeq &-\frac{e^2m^2_\mu}{48\pi^2  c^2_W s^2_W M^2_Z}\, s^2_\zeta ( v^2_\mu -5 a^2_\mu)
\eea
where we took  $M_{Z}\gg m_\mu$ in the second line and $s_\zeta$ is the mixing angle  between $Z$ and $V^0$ gauge bosons, approximately given by $s_\zeta\simeq  -(s_W g_D/g_Y) \sin^2\beta\, (M^2_Z/m^2_V)$ from eq.~(\ref{Zmix}). Thus, we find that the deviation of the SM contribution is positive, and it is given by $ \Delta a^{Z,\mu}_\mu \lesssim 1.8\times 10^{-13}$ for $|s_\zeta|\lesssim 0.01$ or $|\sin\beta|\lesssim 0.3$ with $m_V=500\,{\rm GeV}$ and $v_D=300\,{\rm GeV}$, so we can ignore it in the following discussion.

\subsection{Scalar contributions}

Moreover, $\Delta a^{h,E}_{\mu}$ are the $(g-2)_\mu$ contributions coming from the neutral scalars, $h_i=h, H, s, A^0, {\tilde\varphi}$, running together with the vector-like lepton in loops, given \cite{seesaw,kimiko} by
\bea
\Delta a^{h,E}_\mu &=& \frac{m^2_\mu}{8\pi^2} \sum_{i=1}^5\int^1_0 dx \,
\frac{\Big[|v^E_i|^2 \big(x^2-x^3+\frac{M_E}{m_\mu}\,x^2 \big)+|a^E_i|^2 \{M_E\to -M_E\} \Big]}{m^2_\mu x^2+(M^2_E-m^2_\mu)x+m^2_{h_i}(1-x)} \label{scalarVL}
\eea
where $M_{E'}\simeq M_E$ is taken. 
Then, we get the approximate results for the above contributions, as follows,
\bea
\Delta a^{h,E}_{\mu} \simeq \frac{m^2_{\mu}}{48\pi^2 M^2_{E}}  \Big[ |v^{E}_i|^2+|a^{E}_i|^2+ \frac{3M_{E}}{m_{\mu}}( |v^{E}_i|^2-|a^{E}_i|^2)  \Big] \label{scalar1}
\eea
for $M_{E}\gg m_{h_a}$; 
\bea
\Delta a^{h,E}_{\mu} \simeq  \frac{m^2_{\mu}}{24\pi^2 m^2_{h_i}}  \bigg[  |v^{E}_i|^2+|a^{E}_i|^2+\frac{3M_{E}}{m_{\mu}} ( |v^{E}_i|^2-|a^{E}_i|^2) \Big(\ln\Big(\frac{m^2_{h_i}}{M^2_{E}}\Big) -\frac{3}{2}\Big)\bigg]\label{scalar2}
\eea
for  $M_{E}\ll m_{h_i}$. 

We first consider the $(g-2)_\mu$ contributions from $s$ in eq.~(\ref{scalarVL}).
For  the small mixing angles for the vector-like lepton, we can take  
\bea
|v^{E}_3|^2+|a^{E}_3|^2&\simeq&\frac{1}{4}\Big(\frac{m_{R}}{v_D}\Big)^2, \\
 |v^{E}_3|^2-|a^{E}_3|^2&\simeq &-\frac{1}{2}\Big(\frac{m_{R}}{v_D}\Big)^2 s_{L} s_{R}\simeq  -\frac{1}{2}\Big(\frac{m_{R}}{v_D}\Big)^2 \frac{m_\mu }{M_{E}}.
 \eea
Then, we find that the part containing $|v^{E}_3|^2+|a^{E}_3|^2$ is comparable to the part containing $ |v^{E}_3|^2-|a^{E}_3|^2$, because the latter chirality-enhanced contribution is compensated by the small mixing angles for the vector-like lepton.
Thus, we find that the singlet-like scalar loops with vector-like lepton, $\Delta a^{s,E}_\mu$, contributes a suppressed negative contribution to the muon $g-2$, as shown from eqs~(\ref{scalar1}) and (\ref{scalar2}), and the contributions of the neutral scalars in 2HDM are also negligible in the decoupling limit.

Next we discuss the contribution from $h_5= {\tilde\varphi}$  in eq.~(\ref{scalarVL}), which is the extra contribution in the $SU(2)_D$ case, as compared to the $U(1)'$ case.  For the small mixing angles for the vector-like lepton and $v_2\ll v_D$, we first note that 
\bea
|v^E_5|^2+ |a^E_5|^2&=& \frac{m^2_R}{2v^2_D} \,\sin^2\beta_D\,c^2_R+ \frac{m^2_L}{v^2} \,\frac{\cos^2\beta_D}{\sin^2\beta}\,c^2_L\simeq \frac{m^2_L}{v^2} \,\frac{1}{\sin^2\beta}, \\
 |v^E_5|^2- |a^E_5|^2&= & -\frac{\sqrt{2} m_R m_L}{ v_D v}\, \frac{\sin\beta_D\cos\beta_D}{\sin\beta}\,c_R c_L \simeq -\frac{m_R m_L}{v^2_D}. \label{complexs}
\eea
Thus, the contribution from $h_5= {\tilde\varphi}$ is not suppressed by the mixing angles for the vector-like lepton.
But, the chirality-enhanced contribution proportional to $ |v^E_5|^2- |a^E_5|^2$ turns out to be negative and enhanced by the vector-like lepton mass, as can be seen from eq.~(\ref{scalar1}) or (\ref{scalar2}) with  eq.~(\ref{complexs}).
Thus, for the favored positive correction to $(g-2)_\mu$, we need to take ${\tilde\phi}$ to be sufficiently heavy close to TeV scale. 

On the other hand, $\Delta a^{h,\mu}_\mu$ coming from the neutral scalars and the muon in loops is suppressed further \cite{seesaw}, so we ignore it in the following discussion.

Finally, $\Delta a^{H^-}_\mu$ is the contribution coming from the charged Higgs running in loops \cite{seesaw}, approximated to
\bea
\Delta a^{H^-}_{\mu} &=& \frac{m^2_\mu}{4\pi^2}\,|v_{H^-}|^2\int^1_0 dx \,\frac{x^3-x^2}{m^2_\mu x^2+(m^2_{H^-}-m^2_\mu) x+m^2_\mu(1-x) } \nonumber \\
&\simeq&  -\frac{m^2_{\mu}}{24\pi^2 m^2_{H^-}}\, |v_{H^-}|^2,
\eea
which is doubly suppressed by the mass of the charged Higgs mass and the small mixing angles for the vector-like lepton.

In the left plot of Fig.~\ref{fig:g-2}, we show  the muon $g-2$ correction as a function of the VEV of the dark Higgs doublet, $v_D$, in red solid(dashed) lines for $\sin\theta_R=0.011 (0.033)$ and $\sin\theta_L=0.010$. In the case with $\sin\theta_R=0.011(0.033)$, the seesaw muon mass is equal to (three times larger than) the actual muon mass, so the bare muon mass $m_0$ should be taken to $m_0=2m_\mu (4m_\mu)$ in this case. We fixed the set of the other parameters to $M_E=1\,{\rm TeV}$, $m_{V^+}=500\,{\rm GeV}$, $m_{\tilde\phi}=800\,{\rm GeV}$ and $\sin\beta=0.25$. On the other hand, in the right plot of Fig.~\ref{fig:g-2}, we also present the muon $g-2$ correction as a function of $m_{V^+}$ in red solid(dashed) lines with the same set of the parameters as in the left plot of Fig.~\ref{fig:g-2}, except that $v_D=300\,{\rm GeV}$ and $m_{V^+}$ is allowed to vary.  For comparison, in both plots, the regions favored by the muon $g-2$ within $1\sigma$($2\sigma$) are also shown in yellow(green).

\section{$W$ boson mass}

We show how the anomaly of the $W$ boson mass can be explained in our model, being compatible with the muon $g-2$. 
We focus on the one-loop effects of the vector-like lepton and the tree-level effect of the $Z-V^0$ mass mixing for the $\rho$ parameter.

The theoretical value of the $W$ boson mass can be derived from the muon decay amplitude, which relates $M_W$ to the Fermi constant $G_\mu$, the fine structure constant $\alpha$, and the $Z$ boson mass $M_Z$\cite{Heinemeyer:2004gx}.
In our model, the charged current for the muon is modified  due to the mixing between the muon and the vector-like lepton, so the $W$ boson mass is determined by the experimental inputs with the following modified formula,
\bea
M^2_W \bigg( 1-\frac{M^2_W}{M^2_Z}\bigg) =\frac{\pi \alpha}{\sqrt{2} G_\mu} \,\cos\theta_L \bigg( 1+ \frac{\Delta r}{\cos\theta_L}\bigg) 
\eea
where $\Delta r$ encodes the loop corrections in the SM such as the running $\alpha$, vacuum polarizations for electroweak gauge bosons, and the contributions from new physics. Here, $\cos\theta_L$ is included to account for the modified gauge coupling for the $W$ boson due to the lepton mixing \cite{kimiko}.
For $\Delta r=0.0381$ and $\cos\theta_L=1$ in the SM, we get  the SM prediction for the $W$ boson mass, as follows \cite{Haller:2018nnx,ParticleDataGroup:2020ssz}, 
\bea
M^{\rm SM}_W=80.357\,{\rm GeV}\pm 6\,{\rm MeV}. 
\eea

Recently, the Tevatron CDF II experiment \cite{CDF:2022hxs} has announced  the new measured value of the $W$ boson mass as
\bea
M^{\rm CDFII}_W= 80.4335\,{\rm GeV}\pm 9.4\,{\rm MeV}, \label{dw}
\eea
showing a deviation from the SM prediction at  $7.0\sigma$ level.

\begin{figure}[!t]
\begin{center}
\begin{tikzpicture}[baseline=($0.5*(Win)+0.5*(Wout)$)]
		\begin{feynman}[inline=($0.5*(Win)+0.5*(Wout)$),medium]
			\vertex (Win) at (-2.0, 0) {\( W^- \)};
			\vertex (Wout) at (2.0, 0) {\( W^- \)};
			\vertex (int1) at (-0.7, 0);
			\vertex (int2) at (0.7, 0);
		
			\diagram* {
				(Win) -- 
				[photon] (int1) -- 
				[fermion, half left, edge label = {\( \mu , E \)}] (int2) -- 
				[fermion, half left, edge label = \( \nu \)] (int1),				
				(int2) -- [photon] (Wout),
			};
		\end{feynman}
	\end{tikzpicture}\,\,\,\,\,\,\,\,
\begin{tikzpicture}[baseline=($0.5*(Win)+0.5*(Wout)$)]
		\begin{feynman}[inline=($0.5*(Win)+0.5*(Wout)$),medium]
			\vertex (Win) at (-2.0, 0) {\( W^3 \)};
			\vertex (Wout) at (2.0, 0) {\( W^3 \)};
			\vertex (int1) at (-0.7, 0);
			\vertex (int2) at (0.7, 0);
		
			\diagram* {
				(Win) -- 
				[photon] (int1) -- 
				[fermion, half left, edge label = {\( \mu \)}] (int2) -- 
				[fermion, half left, edge label = {\( \bar \mu \)}] (int1),				
				(int2) -- [photon] (Wout),
			};
		\end{feynman}
	\end{tikzpicture} \\
\begin{tikzpicture}[baseline=($0.5*(Win)+0.5*(Wout)$)]
		\begin{feynman}[inline=($0.5*(Win)+0.5*(Wout)$),medium]
			\vertex (Win) at (-2.0, 0) {\( W^3 \)};
			\vertex (Wout) at (2.0, 0) {\( W^3 \)};
			\vertex (int1) at (-0.7, 0);
			\vertex (int2) at (0.7, 0);
		
			\diagram* {
				(Win) -- 
				[photon] (int1) -- 
				[fermion, half left, edge label = \( E \)] (int2) -- 
				[fermion, half left, edge label = \( \bar E \)] (int1),				
				(int2) -- [photon] (Wout),
			};
		\end{feynman}
	\end{tikzpicture}\,\,\,\,\,\,\,\,
\begin{tikzpicture}[baseline=($0.0*(Win)+0.0*(extH1)$)]
		\begin{feynman}[inline=($0.5*(Win)+0.5*(extH1)$),medium]
			\vertex (Win) at (-2.0, 0) {\( W^3 \)};
			\vertex (Wout) at (2.0, 0) {\( W^3 \)};
			\vertex (int1) at (-0.8, 0);
			\vertex (int2) at (0.8, 0);
			\vertex (extH1) at (-1.4, -1.2) {\( \langle H' \rangle \)};
			\vertex (extH2) at (-0.5, -1.2) {\( \langle H'^\dagger \rangle \)};
			\vertex (extH3) at (0.5, -1.2) {\( \langle H' \rangle \)};
			\vertex (extH4) at (1.4, -1.2) {\( \langle H'^\dagger \rangle \)};
		
			\diagram* {
				(Win) -- 
				[photon] (int1) -- 
				[photon, insertion={[size=4pt, style=thick]0}, edge label = {\( B, V^0 \)}] (int2) -- 
				[photon, insertion={[size=4pt, style=thick]0}] (Wout),
				(extH1) -- [scalar] (int1) -- [scalar] (extH2),
				(extH3) -- [scalar] (int2) -- [scalar] (extH4),
			};
		\end{feynman}
	\end{tikzpicture}			
\end{center}
\caption{Feynman diagrams for the self-energy corrections for electroweak gauge bosons.}
\label{WFeyn}
\end{figure}
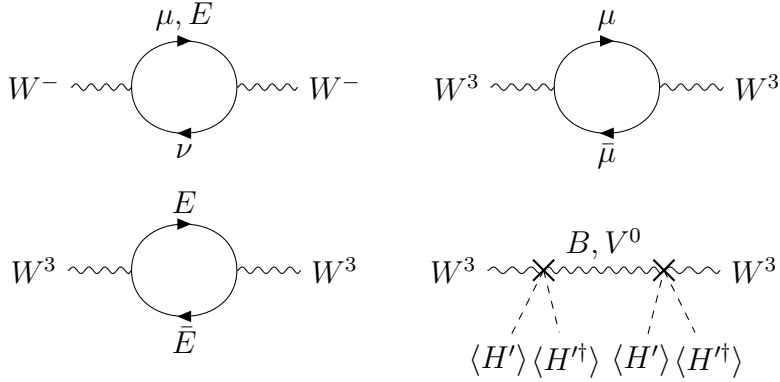

The new physics contribution to the W boson mass can be approximated to
\bea
\Delta M_W \simeq \frac{1}{2} M_W\,\frac{s^2_W}{c^2_W-s^2_W}\, \Big(1-\cos\theta_L+\frac{c^2_W}{s^2_W}\,(\Delta\rho_L+\Delta\rho_H)\Big). \label{Wmass}
\eea
As shown in the loop diagrams (the two plots in the upper panel and the left plot in the lower panel) in Fig.~\ref{WFeyn}, the mixing between the lepton and the vector-like lepton modifies the $\rho$ parameter at one-loop \cite{Lavoura:1992np,seesaw,kimiko} by
\bea
\Delta\rho_L  = \frac{\alpha}{16\pi s^2_W c^2_W}\, \sin^2\theta_L \bigg[\frac{M^2_E}{M^2_Z}-   \frac{m^2_\mu}{M^2_Z}-(\cos^2\theta_L) \theta_+(z_E,z_\mu) \bigg], \label{rhoL}
\eea
with $z_E =M^2_E/M^2_Z$, $z_\mu=m^2_\mu/M^2_Z$, and
\bea
\theta_+(a,b)= a+b -\frac{2ab}{a-b} \ln \frac{a}{b}.
\eea
For $M_E\gtrsim M_Z$, we can make an approximation of the vector-like lepton contribution to the $\rho$ parameter to
\bea
\Delta\rho_L \simeq  \frac{\alpha M^2_E}{16\pi s^2_W c^2_W M^2_Z} \,\sin^4\theta_L.
\eea
We note that there are also one-loop contributions from the vector-like lepton $E'$, ${\tilde\varphi}$ and $V^\pm$, which are proportional to $\sin^2\zeta$, so they are sub-dominant as compared to the tree-level effects of the $Z-V^0$ mass mixing, as will be shortly below. So, we ignore those loop contributions in our analysis.

\begin{figure}[t]
\centering
\includegraphics[width=0.43\textwidth,clip]{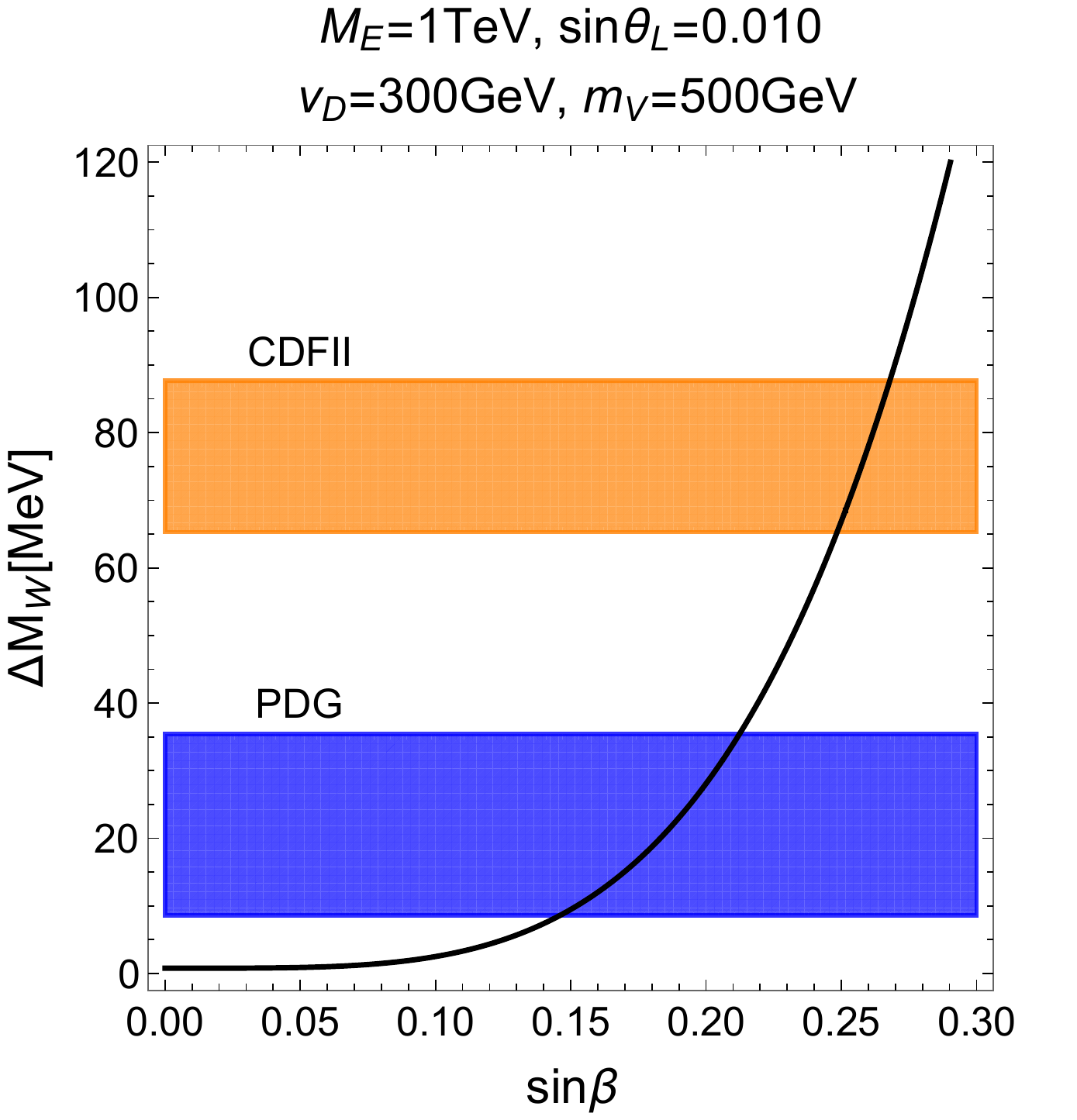}\,\,\,\,\,\,
\includegraphics[width=0.43\textwidth,clip]{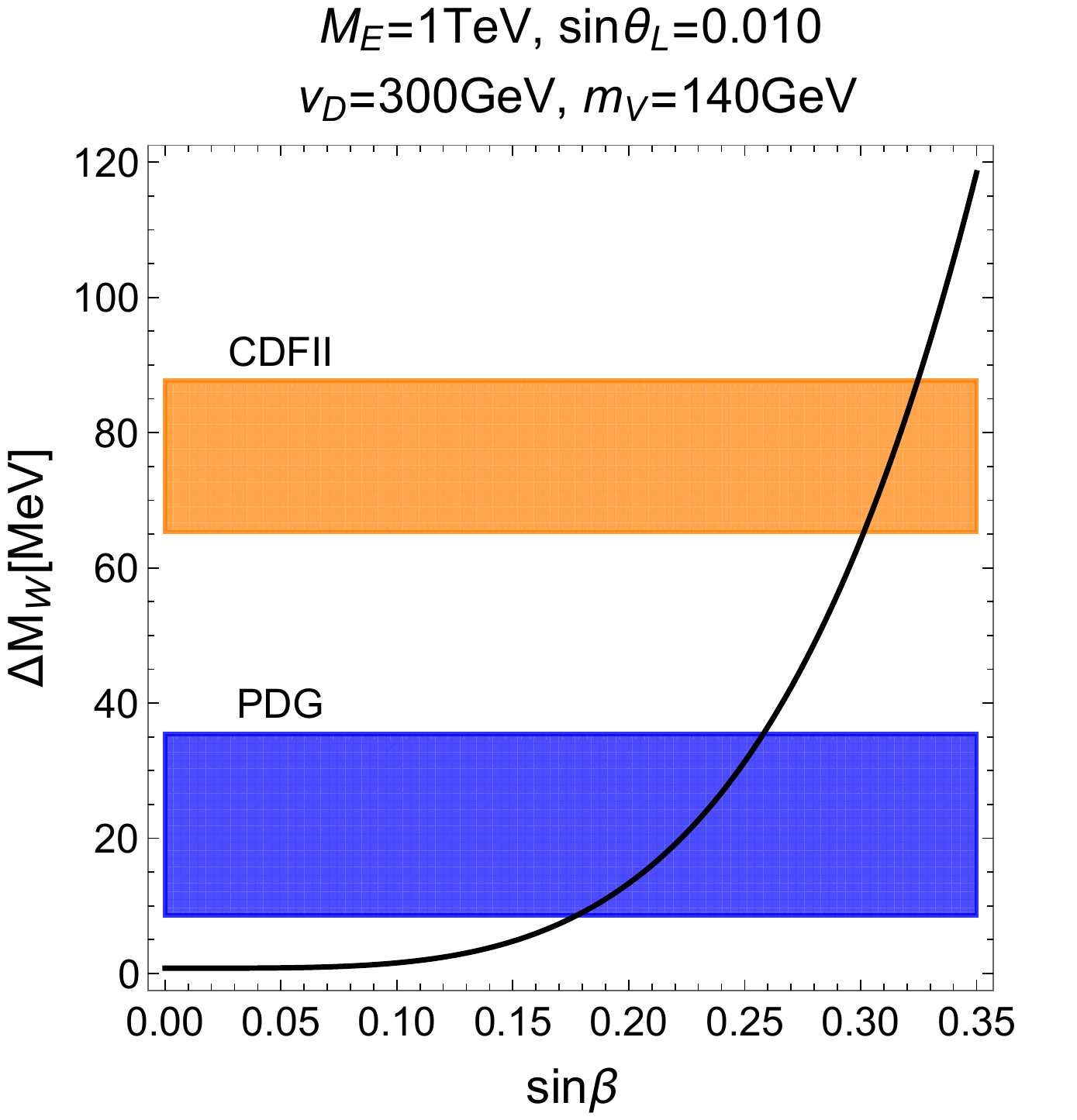}
\caption{$\Delta M_W$  as a function of $\sin\beta$ in black lines.  We took $M_E=1\,{\rm TeV}$, $\sin\theta_L=0.010$ and $v_D=300\,{\rm GeV}$, in common, and  $m_{V}=500(140)\,{\rm GeV}$ on left(right). The deviation of the $W$ boson mass from the SM value is shown within $1\sigma$ errors as inferred from the Tevatron CDFII measurement \cite{CDF:2022hxs}  and the world average of the $W$ boson mass published in Particle Data Group (PDG) \cite{ParticleDataGroup:2020ssz}, in orange and blue regions, respectively.  
}
\label{fig:mw}
\end{figure}

Moreover, as shown in the tree-level diagram (the right plot in the lower panel) in Fig.~\ref{WFeyn}, the VEV of the second Higgs leads to a nonzero mixing between $Z$ and $V^0$ gauge bosons, contributing to the $\rho$ parameter \cite{Bmeson,seesaw,kimiko} by
\bea
\Delta \rho_H =\frac{M^2_W}{M^2_{Z_1} \cos^2\theta_W}\,\cos^2\zeta -1
\eea
where $M_{Z_1}, M_{Z_2}$ are the mass eigenvalues for the $Z$-like and $V^0$-like gauge bosons with $M_{Z_1}<M_{Z_2}$, given in eq.~(\ref{Zmasses}), and $\zeta$ is the mixing angle between $Z$ and $V^0$ gauge bosons, given in eq.~(\ref{Zmix}).
Then, we can approximate the above formula, as follows,
\bea
\Delta \rho_H\simeq  \left\{ \begin{array}{cc} \frac{s^2_W g^2_D}{g^2_Y} \frac{M^2_Z}{m^2_{V^0}}\,\sin^4\beta, \quad  m_{V^0}\gg M_{Z}, \vspace{0.3cm} \\  -\frac{s^2_W g^2_D}{g^2_Y} \,\sin^4\beta, \quad m_{V^0}\ll M_Z. \end{array} \right.
\eea
Here, we note from eq.~(\ref{Zmix}) that for $m_{V^0}\gg M_Z$,  $\sin\beta$ is related to the mass mixing for the neutral gauge bosons by
\bea
\zeta\simeq  -\frac{s_Wg_D}{g_Y} \frac{M^2_Z}{m^2_{V^0}}\,\sin^2\beta.
\eea

In Fig.~\ref{fig:mw}, we show the correction to the $W$ boson mass as a function of $\sin\beta$ in our model with $m_{V^+}=500(140)\,{\rm GeV}$ on left(right). We also took the other parameters to $M_E=1\,{\rm TeV}$, $\sin\theta_L=0.010$, $v_D=300\,{\rm GeV}$, which are consistent with the muon $g-2$ constraint discussed in the previous subsection for Fig.~\ref{fig:g-2}.  We overlaid the regions favored by the Tevatron CDFII measurement \cite{CDF:2022hxs} and the world average in PDG \cite{ParticleDataGroup:2020ssz} for the $W$ boson mass in orange and blue, respectively. In particular, the Tevatron CDFII result requires $\sin\beta=0.25(0.3)$ for $m_{V^+}=500(140)\,{\rm GeV}$ on left(right).

\section{Dark matter constraints}

In this section, we investigate the possibility that a correct relic density for the isospin charged gauge bosons of $SU(2)_L$ is accommodated in our model, being compatible with the muon $g-2$  and the $W$ boson mass and the direct detection bounds on dark matter.

\subsection{Dark matter relic density}

We recall that the $SU(2)_D$ gauge bosons $V^\pm$, the vector-like lepton $E'$ and the neutral complex scalar ${\tilde\varphi}$ are odd under $Z_2$ parity in our model, so the lightest particle among them would be stable.
In order for $V^\pm$ to be a dark matter candidate, we need to take $m_{V^\pm}\simeq m_{V^0}<M_{E'}\simeq M_E, m_{\tilde\varphi}$.  

\begin{figure}[!t]
\begin{center}
\begin{tikzpicture}[baseline=($(int)$)]
		\begin{feynman}[inline=($(int)$),medium]
			\vertex (in1) at (-1.2, 1.2) {\( V^+ \)};
			\vertex (in2) at (-1.2, -1.2) {\( V^- \)};
			\vertex (int) at (0, 0);
			\vertex (out1) at (1.2, 1.2) {\( V^0 \)};
			\vertex (out2) at (1.2, -1.2) {\( V^0 \)};
		
			\diagram* {
				(in1) -- [photon] (int) -- [photon] (out2), 
				(in2) -- [photon] (int) -- [photon] (out1), 
			};
		\end{feynman}
	\end{tikzpicture}\,\,\,\,\,\,\,\,
\begin{tikzpicture}[baseline=($0.5*(int1)+0.5*(int3)$)]
		\begin{feynman}[inline=($0.5*(int1)+0.5*(int3)$),medium]
			\vertex (in1) at (-1.3, 1.4) {\( V^+ \)};
			\vertex (in2) at (-1.3, -1.4) {\( V^- \)};
			\vertex (int1) at (0, 0.6);
			\vertex (int3) at (0, -0.6);
			\vertex (out1) at (1.3, 1.4) {\( V^0 \)};
			\vertex (out2) at (1.3, -1.4) {\( V^0 \)};
		
			\diagram* {
				(in1) -- [photon] (int1) -- [photon] (out1),
				(in2) -- [photon] (int3) -- [photon] (out2),
				(int1) -- [photon, edge label= \( V^+ \)] (int3)
			};
		\end{feynman}
\end{tikzpicture}\,\,\,\,\,\,\,\,
\begin{tikzpicture}[baseline=($(int1)$)]
		\begin{feynman}[inline=($(int1)$),medium]
			\vertex (in1) at (-1.5, 1.2) {\( V^+ \)};
			\vertex (in2) at (-1.5, -1.2) {\( V^- \)};
			\vertex (int1) at (-0.7, 0);
			\vertex (int2) at (0.7, 0);
			\vertex (out1) at (1.5, 1.2) {\( V^0 \)};
			\vertex (out2) at (1.5, -1.2) {\( V^0 \)};
		
			\diagram* {
				(in1) -- [photon] (int1) -- [scalar, edge label' = \( s \)] (int2) -- [photon] (out1), 
				(in2) -- [photon] (int1),
				(int2) -- [photon] (out2),
			};
		\end{feynman}
	\end{tikzpicture}\\
\begin{tikzpicture}[baseline=($(int)$)]
		\begin{feynman}[inline=($(int)$),medium]
			\vertex (in1) at (-1.2, 1.2) {\( V^+ \)};
			\vertex (in2) at (-1.2, -1.2) {\( V^- \)};
			\vertex (int) at (0, 0);
			\vertex (out1) at (1.2, 1.2) {\( s \)};
			\vertex (out2) at (1.2, -1.2) {\( s \)};
		
			\diagram* {
				(in1) -- [photon] (int) -- [scalar] (out2), 
				(in2) -- [photon] (int) -- [scalar] (out1), 
			};
		\end{feynman}
	\end{tikzpicture}\,\,\,\,\,\,\,\,
\begin{tikzpicture}[baseline=($0.5*(int1)+0.5*(int3)$)]
		\begin{feynman}[inline=($0.5*(int1)+0.5*(int3)$),medium]
			\vertex (in1) at (-1.3, 1.4) {\( V^+ \)};
			\vertex (in2) at (-1.3, -1.4) {\( V^- \)};
			\vertex (int1) at (0, 0.6);
			\vertex (int3) at (0, -0.6);
			\vertex (out1) at (1.3, 1.4) {\( s \)};
			\vertex (out2) at (1.3, -1.4) {\( s \)};
		
			\diagram* {
				(in1) -- [photon] (int1) -- [scalar] (out1),
				(in2) -- [photon] (int3) -- [scalar] (out2),
				(int1) -- [photon, edge label = \( V^+ \)] (int3)
			};
		\end{feynman}
	\end{tikzpicture}\,\,\,\,\,\,\,\,
\begin{tikzpicture}[baseline=($(int1)$)]
		\begin{feynman}[inline=($(int1)$),medium]
			\vertex (in1) at (-1.5, 1.2) {\( V^+ \)};
			\vertex (in2) at (-1.5, -1.2) {\( V^- \)};
			\vertex (int1) at (-0.7, 0);
			\vertex (int2) at (0.7, 0);
			\vertex (out1) at (1.5, 1.2) {\( s \)};
			\vertex (out2) at (1.5, -1.2) {\( s \)};
		
			\diagram* {
				(in1) -- [photon] (int1) -- [scalar, edge label' = \( s \)] (int2) -- [scalar] (out1), 
				(in2) -- [photon] (int1),
				(int2) -- [scalar] (out2),
			};
		\end{feynman}
	\end{tikzpicture}\\
\begin{tikzpicture}[baseline=($0.5*(int1)+0.5*(int3)$)]
		\begin{feynman}[inline=($0.5*(int1)+0.5*(int3)$),medium]
			\vertex (in1) at (-1.3, 1.4) {\( V^+ \)};
			\vertex (in2) at (-1.3, -1.4) {\( V^- \)};
			\vertex (int1) at (0, 0.6);
			\vertex (int3) at (0, -0.6);
			\vertex (out1) at (1.3, 1.4) {\( V^0 \)};
			\vertex (out2) at (1.3, -1.4) {\( s \)};
		
			\diagram* {
				(in1) -- [photon] (int1) -- [photon] (out1),
				(in2) -- [photon] (int3) -- [scalar] (out2),
				(int1) -- [photon, edge label = \( V^+ \)] (int3)
			};
		\end{feynman}
	\end{tikzpicture}\,\,\,\,\,\,\,\,	
\begin{tikzpicture}[baseline=($(int1)$)]
		\begin{feynman}[inline=($(int1)$),medium]
			\vertex (in1) at (-1.5, 1.2) {\( V^+ \)};
			\vertex (in2) at (-1.5, -1.2) {\( V^- \)};
			\vertex (int1) at (-0.7, 0);
			\vertex (int2) at (0.7, 0);
			\vertex (out1) at (1.5, 1.2) {\( V^0 \)};
			\vertex (out2) at (1.5, -1.2) {\( s \)};
		
			\diagram* {
				(in1) -- [photon] (int1) -- [photon, edge label' = \( V^0 \)] (int2) -- [photon] (out1), 
				(in2) -- [photon] (int1),
				(int2) -- [scalar] (out2),
			};
		\end{feynman}
	\end{tikzpicture}			
\end{center}
\caption{Dominant processes for dark matter annihilation.}
\label{DMann}
\end{figure}
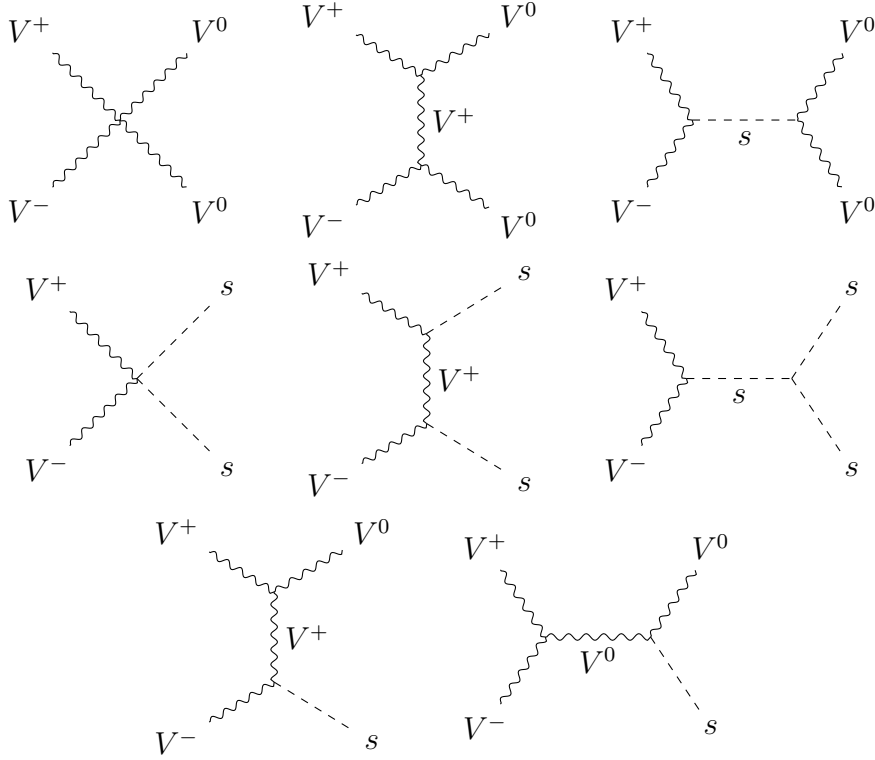

According to the mass spectra obtained in Section 3.2, for a small $\sin\beta$ and $v_D\gtrsim v_2,v$, the scalar fields except the singlet scalar $s$ in our model have common masses, so we can take the decoupling limit of the scalar fields except $s$ to be consistent with the experimental bounds. In this case, dark matter can annihilate dominantly by $V^+V^-\to V^0 s,\, ss$ and $V^+ V^-\to V^0 V^0$, as shown in the Feynman diagrams in Fig.~\ref{DMann}. We note that the $Z-V^0$ mass mixing makes $m_{V^0}$ slightly heavier than $m_{V^\pm}$ as shown in Section 3.1, so the $V^+ V^-\to V^0 V^0$ channel is forbidden at a zero temperature. However, it is kinematically open because of a nonzero velocity of dark matter during the freeze-out \cite{FB,su2d,su2d2}. 

There are other annihilation channels, $V^+ V^-\to hh, V^0 Z$ and $V^+V^-\to {\rm SM}\,{\rm SM}$, but they are suppressed either by $\sin\beta$ or by the mixing angle between the singlet scalar $s$ and the SM Higgs. 
In the following discussion for the dark matter relic density, we ignore the mixing between the singlet scalar $s$ and the SM Higgs and the  $\sin\beta$-induced terms.

The number density $n_{V^+}$ for dark matter $V^+$ is determined by the following Boltzmann equation,
\bea
{\dot n}_{V^+} + 3H n_{V^+} &=& -\langle\sigma v_{\rm rel}\rangle_{V^+ V^-\to V^0 V^0} \Big(n_{V^+} n_{V^-}-n^{\rm eq}_{V^+} n^{\rm eq}_{V^-}\Big) \nonumber \\
&&-\langle\sigma v_{\rm rel}\rangle_{V^+ V^-\to ss} \Big(n_{V^+} n_{V^-}-n^{\rm eq}_{V^+} n^{\rm eq}_{V^-}\Big) \nonumber \\
&&-\langle\sigma v_{\rm rel}\rangle_{V^+ V^-\to V^0 s} \Big(n_{V^+} n_{V^-}-n^{\rm eq}_{V^+} n^{\rm eq}_{V^-}\Big), 
\eea
where $n^{\rm eq}_{V^\pm}$ are the number densities in equilibrium,
and the number density  $n_{V^-}$ for dark matter $V^-$ satisfies the same equation, leading to $n_{V^-}=n_{V^+}$. 
Here, we assumed that $V^0, s$ are in thermal equilibrium with the SM plasma during the freeze-out. 

Then, using the detailed balance condition \cite{su2d,su2d2} for the forbidden channel, $V^+ V^-\to V^0 V^0$, by
\bea
 \langle\sigma v_{\rm rel}\rangle_{V^+ V^-\to V^0 V^0} =\frac{4(n^{\rm eq}_{V^0})^2}{(n^{\rm eq}_{\rm DM})^2} \,\langle\sigma v_{\rm rel} \rangle_{V^0 V^0\to V^+ V^-},
\eea
we obtain the Boltzmann equation for the total dark matter number density, $n_{\rm DM}=n_{V^+}+n_{V^-}=2n_{V^+}$, as follows,
\bea
{\dot n}_{\rm DM} +3H n_{\rm DM} = -\langle\sigma v_{\rm rel}\rangle_{\rm eff}\Big(n^2_{\rm DM}-(n^{\rm eq}_{\rm DM})^2 \Big)
+\langle\sigma v_{\rm rel}\rangle_{\rm FB} \, (n^{\rm eq}_{V^0})^2 \bigg(1-\frac{n^2_{\rm DM}}{(n^{\rm eq}_{\rm DM})^2}\bigg)
\label{Boltzmann}
\eea
with
\bea
\langle\sigma v_{\rm rel}\rangle_{\rm eff}&=&\frac{1}{2}\langle\sigma v_{\rm rel}\rangle_{V^+ V^-\to ss}+\frac{1}{2}\langle\sigma v_{\rm rel}\rangle_{V^+ V^-\to V^0 s}, \\
\langle\sigma v_{\rm rel}\rangle_{\rm FB} &=&2\langle\sigma v_{\rm rel}\rangle_{V^0 V^0\to V^+ V^-}.
\eea
Here, in the non-relativistic limit for dark matter, the annihilation cross sections are given by
\bea
\langle\sigma v_{\rm rel}\rangle_{V^+ V^-\to ss}&=& \frac{|{\cal M}_{V^+ V^-\to ss}|^2}{32\pi m^2_{V^+}}
\sqrt{1-\frac{m^2_s}{m^2_{V^+}}}, \\
\langle\sigma v_{\rm rel}\rangle_{V^+ V^-\to V^0 s}&=& \frac{|{\cal M}_{V^+ V^-\to V^0 s}|^2}{32\pi m^2_{V^+}}
\sqrt{1-\frac{(m_s+m_{V^0})^2}{4m^2_{V^+}}}\sqrt{1-\frac{(m_s-m_{V^0})^2}{4m^2_{V^+}}}, \\
\langle\sigma v_{\rm rel}\rangle_{V^0 V^0\to V^+ V^-}&=& \frac{|{\cal M}_{V^0 V^0\to V^+ V^-}|^2}{64\pi m^2_{V^0}}\sqrt{1-\frac{m^2_{V^+}}{m^2_{V^0}}},
\eea
with
\bea
|{\cal M}_{V^+ V^-\to ss}|^2 &=&\frac{g^4_D}{72}\bigg( \frac{4g^4_D v^4_D}{m^4_{V^+}(m^2_s-2m^2_{V^+})^2}\,(m^4_s-4m^2_s m^2_{V^+}+6m^4_{V^+} ) \nonumber \\
&&+\frac{4g^2_D v^2_D}{m^2_{V^+}(2m^2_{V^+}-m^2_s )}\,(m^2_s-4m^2_{V^+}-12 \lambda_\phi v^2_D) \nonumber \\
&&+\frac{3}{(m^2_s-4m^2_{V^+})^2} \,(m^2_s-4m^2_{V^+}-12\lambda_\phi v^2_D )^2   \bigg),
\eea
\bea
|{\cal M}_{V^+ V^-\to V^0 s}|^2 &=&\frac{g^6_D v^2_D(4m^2_{V^+}-m^2_{V^0})^2}{18 m^6_{V^+}(m^2_s-4m^2_{V^+}+m^2_{V^0})^2}\,\times\nonumber\\
&&\times \Big(m^4_s+(4m^2_{V^+}-m^2_{V^0})^2-2m^2_s (4m^2_{V^+}+m^2_{V^0}) \Big),
\eea
and
\bea
|{\cal M}_{V^0 V^0\to V^+ V^-}|^2 &=&\frac{4g^4_D}{9 m^8_{V^+} m^4_{V^0}} \,\Big( 48m^{12}_{V^+} +88 m^{10}_{V^+} m^2_{V^0}-71 m^8_{V^+}m^4_{V^0}+64 m^6_{V^+} m^6_{V^0} \nonumber \\
&&\quad+m^4_{V^+}m^8_{V^0} -2m^2_{V^+}m^{10}_{V^0}+ m^{12}_{V^0}    \Big) \nonumber \\
&&-\frac{2g^6_D v^2_D }{9m^6_{V^+} m^2_{V^0}(m^2_s-4m^2_{V^0}) }\, \Big(12m^8_{V^+}-23 m^6_{V^+}m^2_{V^0} +10 m^4_{V^+}m^4_{V^0} \nonumber \\
&&\quad-3m^2_{V^+}m^6_{V^0} -2m^8_{V^0}  \Big) \nonumber \\
&&+\frac{g^8_D v^4_D }{12m^4_{V^+}(m^2_s-4m^2_{V^0})^2} \, (3m^4_{V^+}-4m^2_{V^+} m^2_{V^0} +4m^4_{V^0}).
 \eea

\begin{figure}[t]
\centering
\includegraphics[width=0.43\textwidth,clip]{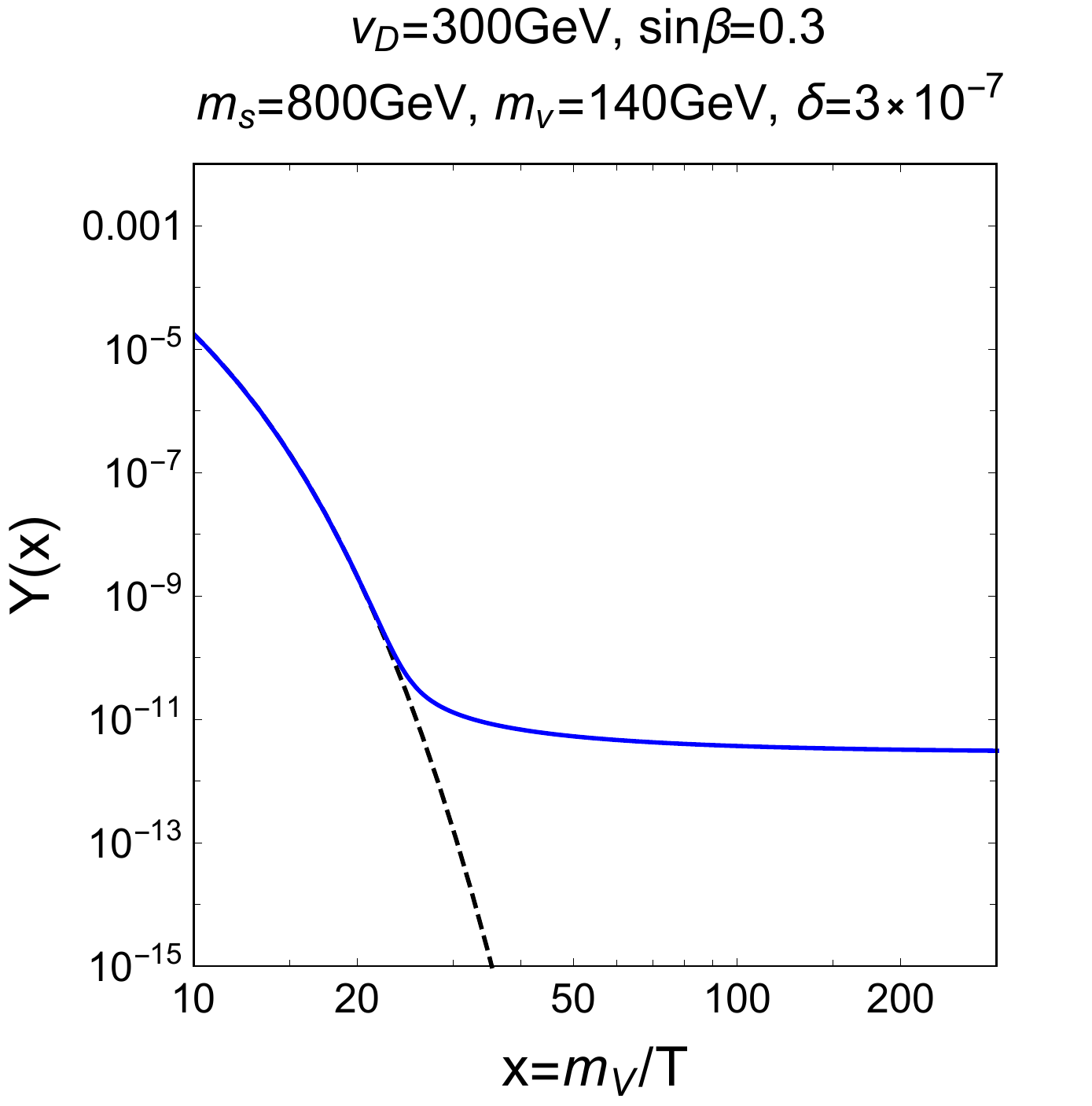}\,\,\,\,\,\,
\includegraphics[width=0.43\textwidth,clip]{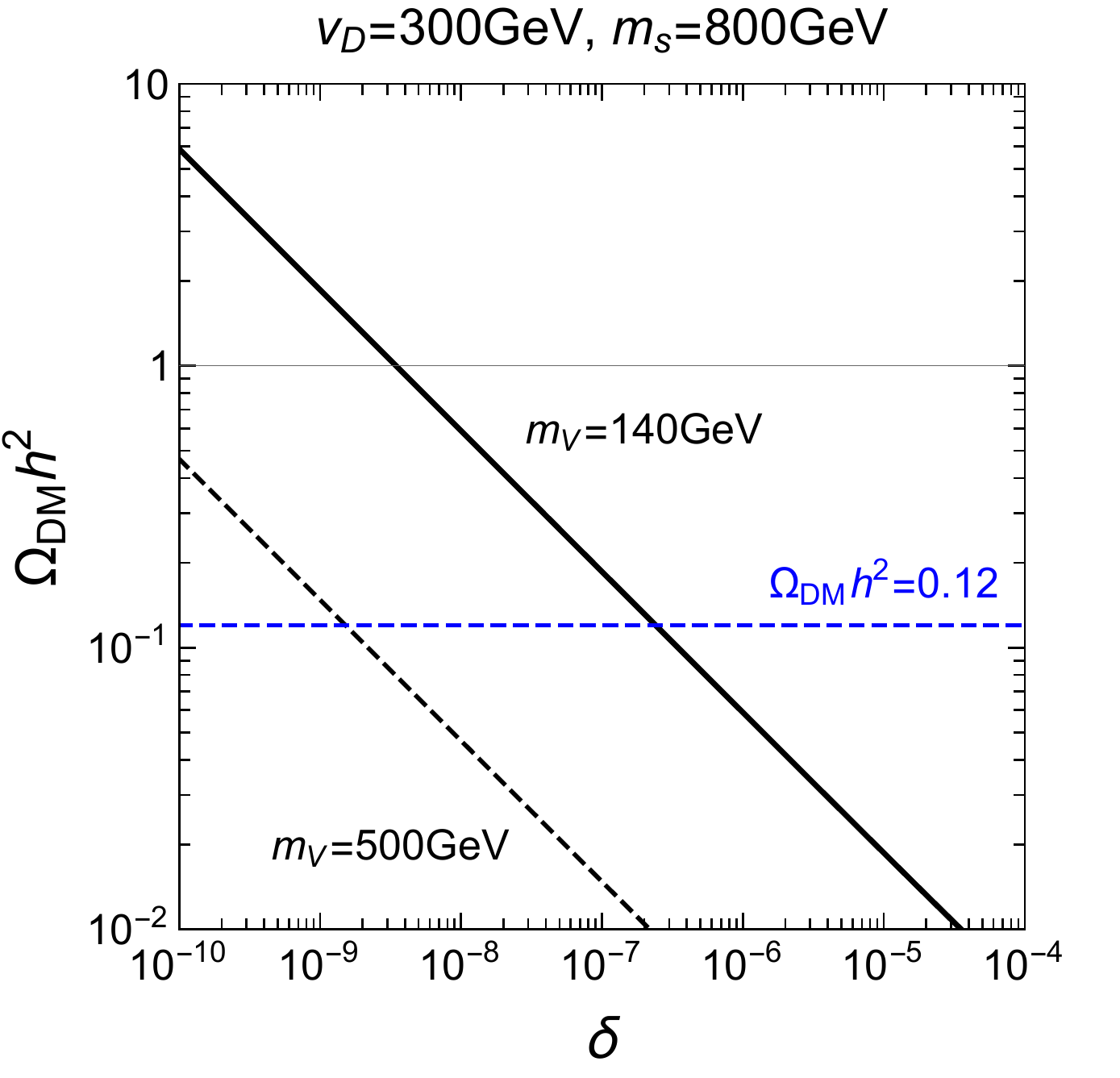}
\caption{(Left) The relic abundance for dark matter as a function of $x=m_{V^+}/T$. (Right) Dark matter relic density as a function of $\delta\equiv m_{V^0}/m_{V^+}-1$. For both plots, we took $v_D=300\,{\rm GeV}$ and $m_s=800\,{\rm GeV}$. We also chose $m_{V^+}=140\,{\rm GeV}$, $\sin\beta=0.3$ and $\delta=3\times 10^{-7}$ on left, and $\sin\beta=0.3 (0.25)$ and $m_{V^+}=140(500)$ GeV in solid and dashed black lines on right. 
}
\label{fig:relic}
\end{figure}

After solving the Boltzmann equation in eq.~(\ref{Boltzmann}), we can obtain the relic density for dark matter as follows,
\bea 
\Omega_{\rm DM} h^2= 0.2745 \bigg(\frac{Y_{\rm DM}}{10^{-11}} \bigg)\bigg(\frac{m_{V^+}}{100\,{\rm GeV}} \bigg)
\eea
where $Y_{\rm DM}= n_{\rm DM}/s$ is the abundance at present. Here, in the case where $V^+V^-\to V^0 s\,, ss$ channels are kinematically closed for a heavy singlet scalar, which is favored for the consistency with direct detection, we get the analytic expression for the relic abundance at present \cite{su2d,su2d2} by
\bea
Y_{\rm DM}= \frac{x_f H(m_{V^+})}{s(m_{V^+})}\, e^{2\delta x_f} \,f(\delta,x_f)
\eea
where $H(m_{V^+})$ and $s(m_{V^+})$ are the entropy density and the Hubble parameter evaluated at $T=m_{V^+}$, respectively, $x_f=m_{V^+}/T_f$ with $T_f$ being the freeze-out temperature, and
\bea
f(\delta,x_f)=\bigg[\frac{1}{2}\langle\sigma v_{\rm rel}\rangle_{V^0 V^0\to V^+ V^-} (1+\delta)^3 \Big(1-2\delta\, x_f \, e^{2\delta x_f} \int^\infty_{2\delta x_f} dt t^{-1} e^{-t} \Big) \bigg]^{-1}
\eea
with $\delta\equiv (m_{V^0}-m_{V^+})/m_{V^+}$.

In Fig.~\ref{fig:relic}, we depict the relic abundance for dark matter as a function of $x=m_{V^+}/T$ in blue line on left, and the dark matter relic density as a function of $\delta\equiv m_{V^0}/m_{V^+}-1$ for $m_{V^+}=140(500)\,{\rm GeV}$ and $\sin\beta=0.3 (0.25)$ in black solid(dashed) lines, on right. 
We took $v_D=300\,{\rm GeV}$ and $m_s=800\,{\rm GeV}$ in common for both plots, and $m_{V^+}=140\,{\rm GeV}$, $\sin\beta=0.3$ and $\delta=3\times 10^{-7}$ on left.  Our parameter choice is consistent with both the muon $g-2$ and the $W$ boson mass, discussed in the previous subsections. Here, we took a relatively large $m_s$ because of the XENON1T bound, as will be discussed in the next subsection. 
We find that the condition for a correct relic density \cite{planck} is insensitive to $m_s$ and $\sin\beta$ as far as $m_s\gtrsim m_{V^+}$ and $v\sin\beta\lesssim v_D$, but it depends crucially on the mass splitting $\delta$ and the dark matter mass $m_{V^+}$ for a given $v_D$. 
For a fixed $v_D$ and $\sin\beta$, the larger the DM mass, the larger the $SU(2)_D$ gauge coupling, leading to a larger annihilation cross section and a smaller relic abundance for dark matter, as shown in the right plot of Fig.~\ref{fig:relic}.

We remark that the forbidden channel, $V^+ V^-\to V^0 V^0$, is closed if the relative velocity between dark matter particles is sufficiently small, namely, $v_{\rm rel}\lesssim \sqrt{8\delta}$. So, for $v_{\rm rel}\simeq 220\,{\rm km}$ in our galaxy, the forbidden channel does not lead to observable signatures for $\delta\gtrsim 6\times 10^{-7}$.
But, the other subdominant channels for the dark matter relic density, such as $V^+ V^-\to hh, V^0 Z$ and $V^+V^-\to {\rm SM}\,{\rm SM}$, could lead to interesting signals for indirect detection experiments such as Cosmic Microwave Background or cosmic rays \cite{fmdm,srdm}.

\subsection{Dark matter direct detection}

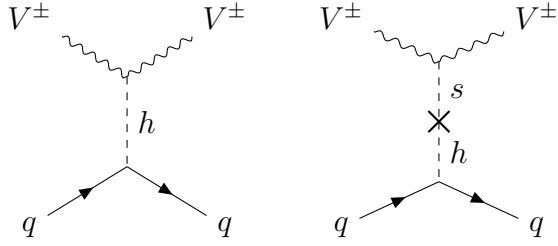
\begin{figure}[!t]
\begin{center}
\begin{tikzpicture}[baseline=($0.5*(int1)+0.5*(int3)$)]
		\begin{feynman}[inline=($0.5*(int1)+0.5*(int3)$),medium]
			\vertex (in1) at (-1.3, 1.4) {\( V^\pm \)};
			\vertex (in2) at (-1.3, -1.4) {\( q \)};
			\vertex (int1) at (0, 0.6);
			\vertex (int3) at (0, -0.6);
			\vertex (out1) at (1.3, 1.4) {\( V^\pm \)};
			\vertex (out2) at (1.3, -1.4) {\( q \)};
		
			\diagram* {
				(in1) -- [photon] (int1) -- [photon] (out1),
				(in2) -- [fermion] (int3) -- [fermion] (out2),
				(int1) -- [scalar, edge label = \( h \)] (int3)
			};
		\end{feynman}
	\end{tikzpicture}\,\,\,\,\,\,\,\,
\begin{tikzpicture}[baseline=($0.5*(int1)+0.5*(int3)$)]
		\begin{feynman}[inline=($0.5*(int1)+0.5*(int3)$),medium]
			\vertex (in1) at (-1.3, 1.4) {\( V^\pm \)};
			\vertex (in2) at (-1.3, -1.4) {\( q \)};
			\vertex (int1) at (0, 0.8);
			\vertex (int2) at (0, 0);
			\vertex (int3) at (0, -0.8);
			\vertex (out1) at (1.3, 1.4) {\( V^\pm \)};
			\vertex (out2) at (1.3, -1.4) {\( q \)};
		
			\diagram* {
				(in1) -- [photon] (int1) -- [photon] (out1),
				(in2) -- [fermion] (int3) -- [fermion] (out2),
				(int1) -- [scalar, edge label = \( s \)] (int2) -- [scalar, insertion={[size=4pt, style=thick]0}, edge label = \( h \)] (int3)
			};
		\end{feynman}
	\end{tikzpicture}	
\end{center}
\caption{Elastic scattering processes for dark matter direct detection.}
\label{DD}
\end{figure}

The annihilation of dark matter into a quark pair, $V^+V^-\to q{\bar q}$, is possible with SM Higgs and singlet scalar exchanges, but it is subdominant for determining the dark matter relic density. Nonetheless, the same interaction can be constrained by the direct detection experiments such as XENON1T \cite{xenon1t}. 

For the spin-independent elastic scattering between dark matter and nucleons, we first get the effective Lagrangian for dark matter and quarks, due to the Higgs and singlet scalar exchanges, shown in Fig.~\ref{DD}, as
\bea
{\cal L}_{V^\pm-q} = \lambda_{\rm eff} m_q V^+_\mu V^{-\mu} {\bar q}q. 
\eea
where
\bea
\lambda_{\rm eff}&=& \frac{\sqrt{2}}{2v}\, v_D g^2_D\sin\theta_ h\cos\theta_h \bigg(\frac{1}{m^2_s}-\frac{1}{m^2_h} \bigg) \nonumber \\
&&-\frac{1}{2}  g^2_D\sin^2\beta\, \bigg(\frac{\sin^2\theta_h}{m^2_s}+\frac{\cos^2\theta_h}{m^2_h} \bigg). \label{lambeff}
\eea
Here, we included the mixing angle $\theta_h$  between the SM Higgs and the singlet scalar, and we used the same notations for the mass eigenstates for Higgs-like scalars, $h, s$, for convenience.

As a result, the corresponding cross section for the elastic scattering between dark matter and nucleon \cite{VDM,LQDM} is given by
\bea
\sigma^{\rm SI}_{V-N} = \frac{\mu^2_{N}}{4\pi m^2_{V^+} A^2} \Big( Z f_p+ (A-Z) f_n \Big)^2
\eea 
where $Z, A-Z$ are the number of protons and neutrons in the detector nucleon, $\mu_N=m_N m_{V^+}/(m_N+m_{V^+})$ is the reduced mass of DM-nucleus system, and 
\bea
f_{p,n} =m_{p,n} \lambda_{\rm eff} \Big(\sum_{q=u,d,s} f^{p,n}_{Tq}+ \frac{2}{9} f^{p,n}_{TG} \Big)
\eea
with $f^{p,n}_{TG}=1-\sum_{q=u,d,s} f^{p,n}_{Tq}$.
 Here, $f^N_{Tq}$ is the mass fraction of quark $q$ inside the nucleon $N$, defined by $\langle N|m_q {\bar q}q |N\rangle= m_N f^N_{Tq}$, and $f^{N}_{TG}$ is the mass fraction of gluon $G$  the nucleon $N$, due to heavy quarks \cite{hisano}. We quote the updated numerical values as $f^p_{T_u}=0.0208\pm 0.0015$ and $f^p_{T_d}=0.0411\pm 0.0028$ for a proton, $f^n_{T_u}=0.0189\pm 0.0014$ and $f^n_{T_d}=0.0451\pm 0.0027$ for a neutron \cite{DDupdate}, and  $f^{p,n}_{T_s}=0.043\pm 0.011$ for both proton and neutron \cite{DDstrange}.

\begin{figure}[!t]
\centering
\includegraphics[width=0.43\textwidth,clip]{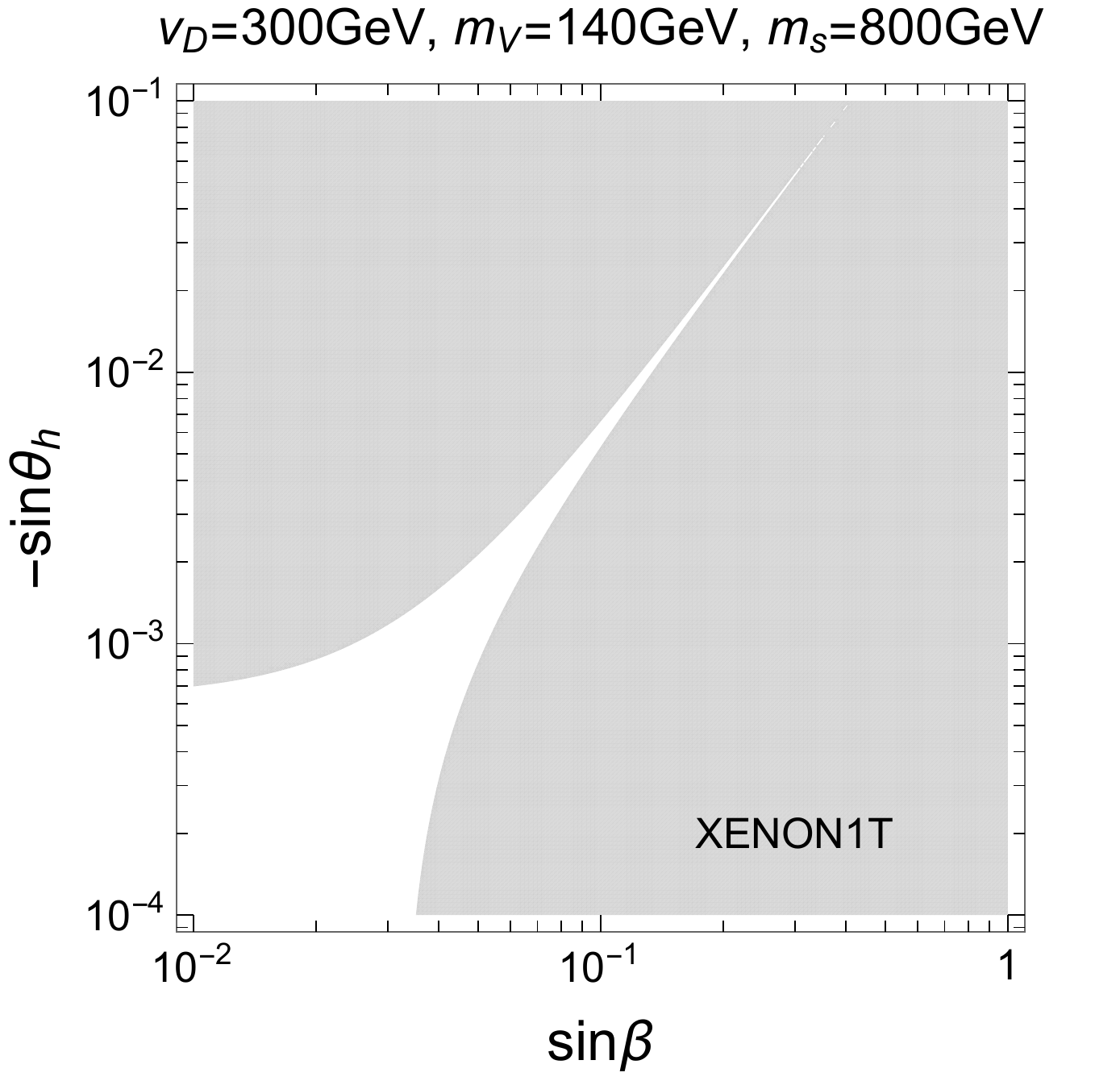}\,\,\,\,\,\,
\includegraphics[width=0.43\textwidth,clip]{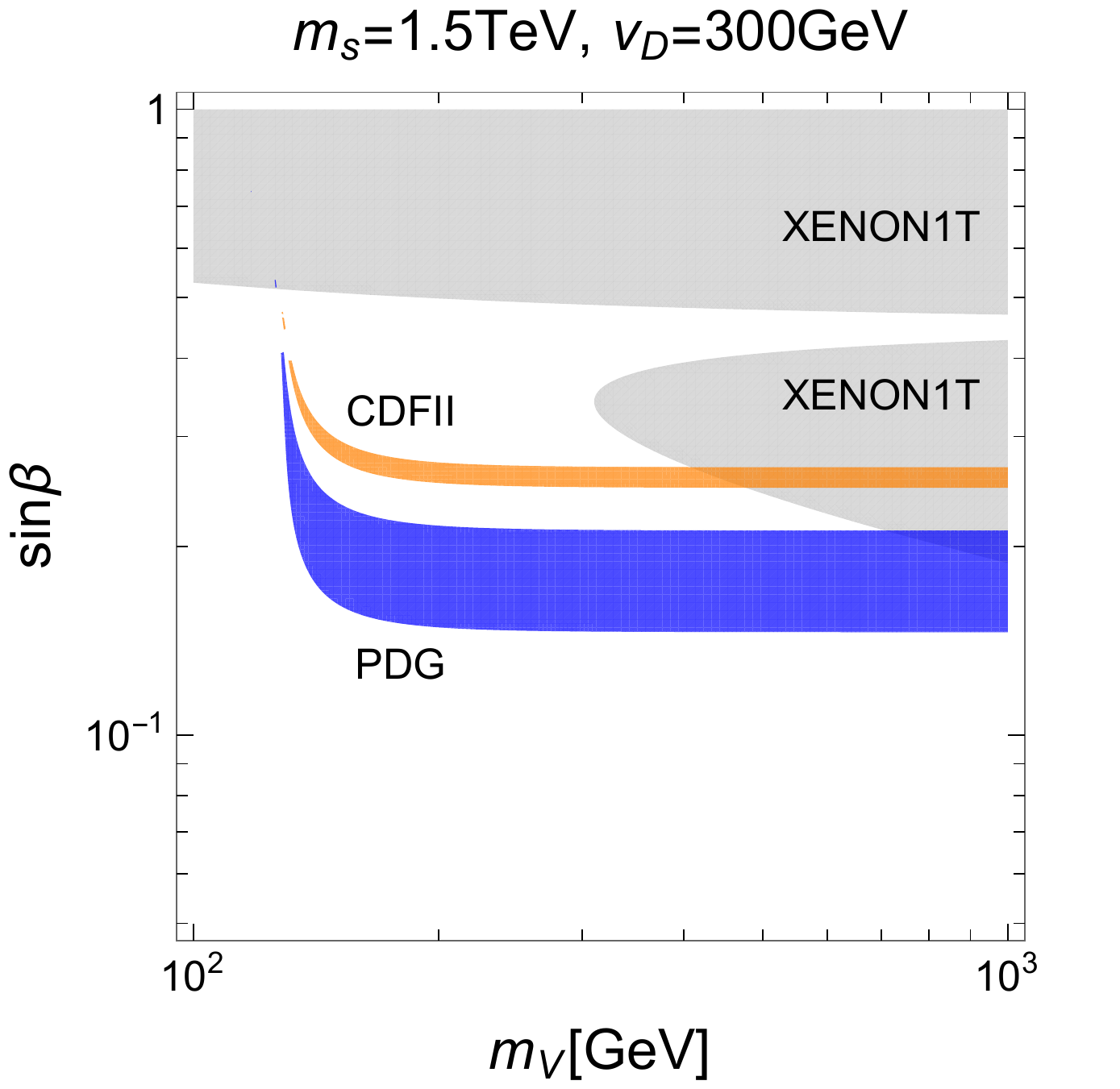}
\caption{(Left) Parameter space in $\sin\beta$ vs $-\sin\theta_h$, that is consistent with  XENON1T \cite{xenon1t}. 
We took $v_D=300\,{\rm GeV}$, $m_{V^+}=140\,{\rm GeV}$ and $m_s=800\,{\rm GeV}$.  (Right) Parameter space in $m_{V^+}$ vs $\sin\beta$, that is consistent with XENON1T.  We took $m_s=1.5\,{\rm TeV}$, $v_D=300\,{\rm GeV}$ and the alignment limit for the Higgs mixing angle by $\sin\theta_h= -\frac{v}{\sqrt{2}v_D} \,\sin^2\beta$. The deviation of the $W$ boson mass from the SM value is shown within $1\sigma$ errors as inferred from the Tevatron CDFII measurement  \cite{CDF:2022hxs} and the world average in PDG \cite{ParticleDataGroup:2020ssz}, in orange and blue regions, respectively.  The gray regions in both plots are ruled out by XENON1T.
}
\label{fig:dd}
\end{figure}

Consequently, we find that the XENON1T bound \cite{xenon1t} on the DM-nucleon scattering cross section can be satisfied in the alignment limit for the Higgs mixing angle, satisfying $\sin\theta_h\simeq -\frac{v}{\sqrt{2}v_D} \,\sin^2\beta$ and $m_s\gg m_h$, for which there is a cancellation  for the Higgs exchanges in the scattering amplitude in eq.~(\ref{lambeff}). 
However, a relatively light singlet scalar belonging to the $SU(2)_D$ doublet can contribute to the DM-nucleon scattering cross section, so dark matter can be tested in the future DM experiments. 

In the left plot of Fig.~\ref{fig:dd}, we show the gray region of the parameter space for $\sin\beta$ and $-\sin\theta_h$, which is ruled out by the XENON1T bound \cite{xenon1t}. Here, we chose $v_D=300\,{\rm GeV}$, $m_{V^+}=140\,{\rm GeV}$ and $m_s=800\,{\rm GeV}$. 
In the right plot of Fig.~\ref{fig:dd}, we also present the gray region of the parameter space for $m_V$ vs $\sin\beta$, which is incompatible with  the XENON1T bound. In the same plot, the orange and blue regions are favored by the Fermilab CDFII measurement and the world average in PDG, respectively, and they are consistent with the XENON1T bound, due to the alignment limit with  $\sin\theta_h\simeq -\frac{v}{\sqrt{2}v_D} \,\sin^2\beta$ and a heavy singlet scalar with $m_s=1.5\,{\rm TeV}$.
Therefore, from the results in Fig.~\ref{fig:g-2} and \ref{fig:mw} and the right plots in Figs.~\ref{fig:relic} and \ref{fig:dd}, we find that there is a consistent parameter space explaining the muon $g-2$ and the the $W$ boson mass as well as dark matter constraints at the same time.

\section{Conclusions}

We considered a new model for lepton flavor and dark matter by extending the SM with an $SU(2)_D$ gauge symmetry and  vector-like leptons in its fundamental representation. The flavor mixing between the lepton and the vector-like lepton is realized by the VEVs of the $SU(2)_D$ Higgs doublet and the Higgs bi-doublet under $SU(2)_D\times SU(2)_L$, giving rise to the seesaw lepton masses.  

Vector-like leptons and $SU(2)_D$ gauge bosons contribute dominantly to the muon $g-2$, being insensitive to the masses of the vector-like leptons, as far as $SU(2)_D$ gauge bosons are relatively light, because a large violation of chirality from the vector-like lepton masses is compensated by the small mixing angles of the muon. 
The vector-like leptons also make corrections to the $W$ boson mass at one-loop, although suppressed due to a small lepton mixing.  However, the mass mixing between the $Z$ boson and the dark isospin-neutral gauge boson ($V^0$)  of $SU(2)_D$ accounts for the deviation of the $W$ boson mass at tree level, as observed in the recent Tevatron CDFII measurement.

A combination of the $U(1)_G$ global symmetry in the Higgs sector and the dark isospin symmetry leads to a $Z_2$ parity, allowing for the isospin-charged gauge boson ($V^\pm$) of $SU(2)_D$ to be dark matter candidates. $SU(2)_D$ gauge boson masses are almost degenerate thanks to the dark custodial symmetry, but a small positive correction to the vector dark matter appears in proportion to the $Z-V^0$ mixing contribution to the $\rho$ parameter.  Then, we also showed that the dark matter constraints such as the relic density and the direct detection bounds can be satisfied in our model, being compatible with the solutions to the anomalies in the muon $g-2$ and the $W$ boson mass. The forbidden annihilation channel for dark matter, $V^+ V^-\to V^0 V^0$, is responsible for explaining the correct relic density whereas the direct detection bounds for dark matter can be satisfied in the alignment limit of the mixing between the SM Higgs and the singlet scalar of the  $SU(2)_D$ doublet.

\section*{Acknowledgments}

The work is supported in part by Basic Science Research Program through the National
Research Foundation of Korea (NRF) funded by the Ministry of Education, Science and
Technology (NRF-2022R1A2C2003567 and NRF-2021R1A4A2001897). 
The work of KY is supported by Brain Pool program funded by the Ministry of Science and ICT through the National Research Foundation of Korea(NRF-2021H1D3A2A02038697).

\begin{appendix}

\section{Mass matrices for scalar and gauge bosons}

The mass matrix for CP-even scalars $(\rho_1,\rho_2,s)$ is given by
\bea
{\cal M}^2_S =\left(\begin{array}{ccc} 2\lambda_1 v^2_1+\frac{\mu_3 v_2v_D}{v_1} & \lambda_3v_1 v_2-\mu_3 v_D &\sqrt{2}\lambda_{H\Phi}v_1 v_D-\frac{1}{\sqrt{2}} \mu_3 v_2  \vspace{0.2cm} \\ \lambda_3v_1 v_2-\mu_3 v_D & 2\lambda_2 v^2_2+\frac{\mu_3 v_1 v_D}{v_2} &\sqrt{2}\lambda_{H'\Phi}v_2 v_D-\frac{1}{\sqrt{2}} \mu_3 v_1 \vspace{0.2cm}   \\ \sqrt{2}\lambda_{H\Phi}v_1 v_D-\frac{1}{\sqrt{2}} \mu_3 v_2 &\sqrt{2}\lambda_{H'\Phi}v_2 v_D-\frac{1}{\sqrt{2}} \mu_3 v_1 &  4\lambda_\phi v^2_D+\frac{\mu_3 v_1 v_2}{2v_D} \end{array} \right).
\eea
Here, we have used the minimization conditions in eqs.~(\ref{min1}), (\ref{min2}) and (\ref{min3}).
The above mass matrix can be diagonalized by
\bea
\left(\begin{array}{c}\rho_1 \\ \rho_2 \\s  \end{array}\right)  = R_S \left( \begin{array}{c} h_1 \\ h_2 \\  h_3 \end{array}\right) 
\eea
where $R_S$ is the $3\times 3$ rotation matrix \cite{Bmeson}. Then, the mass eigenvalues are denoted as $m^2_{h_i}$ with $i=1,2,3$. 
Ignoring the mixing with the singlet scalar $s$, we can simply diagonalize the mass matrix for $\rho_1, \rho_2$ by 
\bea
\left(\begin{array}{c}\rho_1 \\ \rho_2  \end{array}\right)  = \left( \begin{array}{cc} \cos\alpha & -\sin\alpha \\ \sin\alpha & \cos\alpha  \end{array}\right)  \left( \begin{array}{c} h \\ H \ \end{array}\right). 
\eea  
As discussed in the text, $\alpha=\beta$ is the alignment limit, for which we can identify $h$ with the SM Higgs. For our discussion on more general cases,  we can also introduce a mixing angle  $\theta_h$ between the Higgs-like scalar $h$ and the singlet scalar $s$ by
\bea
\left(\begin{array}{c} h \\ s  \end{array}\right)  = \left( \begin{array}{cc} \cos\theta_h & \sin\theta_h \\ -\sin\theta_h   & \cos\theta_h \end{array}\right)  \left( \begin{array}{c} h_1 \\ h_3 \ \end{array}\right).
\eea

The mass matrix for CP-odd scalars $(\eta_1,\eta_2, a)$ is given by
\bea
{\cal M}^2_P= \mu_3 \left(\begin{array}{ccc} \frac{v_2v_D}{v_1}   & -v_D & -\frac{1}{\sqrt{2}}v_2 \vspace{0.2cm} \\ -v_D &\frac{v_Dv_1}{v_2} &  \frac{1}{\sqrt{2}}v_1  \vspace{0.2cm} \\  -\frac{1}{\sqrt{2}}v_2 &  \frac{1}{\sqrt{2}}v_1 & \frac{v_1v_2}{2v_D} \end{array} \right).
\eea
Then, the above mass matrix is diagonalized by
\bea
\left(\begin{array}{c}\eta_1 \\ \eta_2 \\a  \end{array}\right)  =R_P \left( \begin{array}{c} G_Y \\ G^0_D \\ A^0 \end{array}\right) 
\eea
with
\bea
R_P&=& \left(\begin{array}{ccc} \frac{v_1}{v} & \frac{v_2}{v} & 0  \vspace{0.2cm} \\ 0 & -\frac{v_2}{C} & \frac{\sqrt{2} v_D}{C}\vspace{0.2cm}  \\ \frac{N v_2}{v}  & -\frac{Nv_1}{v} & - \frac{Nv_1 v_2}{\sqrt{2} v v_D}\end{array} \right)^{-1} \nonumber \\
&=&  \left(\begin{array}{ccc} \frac{ N^2 C^2v_1}{2v v^2_D } & \frac{ N^2 Cv_1 v^2_2}{2 v^2v^2_D} & \frac{N  v_2 }{v} \vspace{0.2cm} \\ \frac{N^2 v_2}{v }  & -\frac{N^2 Cv^2_1 v_2}{2v^2 v^2_D } & -\frac{N v_1 }{v}\vspace{0.2cm}  \\  \frac{ N^2 v^2_2}{\sqrt{2}v v_D}  & \frac{N^2C}{\sqrt{2}v_D} & -\frac{N v_1v_2}{\sqrt{2}v v_D} \end{array} \right)
\eea
where  the normalization factors are
\bea
C= \sqrt{v^2_2+2v^2_D}, \quad N = \frac{1}{\sqrt{1+\frac{v^2_1 v^2_2}{2v^2 v^2_D}}}.
\eea
There are two massless scalars corresponding to $G_Y, G^0_D$, and the mass of the CP-odd scalar $A^0$ is given in eq.~(\ref{CPoddmass}). 

The mass matrix for charged scalars $(\phi^+_1, \phi^+_2)$ is given by
\bea
{\cal M}^2_C =\mu_3\left(\begin{array}{cc}  \frac{v_2 v_D}{v_1} & -v_D \vspace{0.2cm} \\ -v_D & \frac{v_1 v_D}{v_2} \end{array} \right).
\eea
Then, the above mass matrix is diagonalized by
\bea
\left(\begin{array}{c}\phi^+_1  \\ \phi^+_2  \end{array}\right)  =  \left( \begin{array}{cc} \cos\beta & \sin\beta \\ \sin\beta & -\cos\beta  \end{array}\right)  \left( \begin{array}{c} G^+ \\ H^+  \end{array}\right). 
\eea
There is one massless charged scalar correspond to $G^+$, and the mass of the charged scalar is given by eq.~(\ref{Cmass1}).

The mass matrix for dark complex scalars $(\varphi_1, ({\hat \phi}^0_2)^*)$ is given by
\bea
{\cal M}^2_N=\mu_3\left(\begin{array}{cc}  \frac{v_1 v_2}{2v_D} & -\frac{1}{\sqrt{2}}v_1 \vspace{0.2cm}  \\   -\frac{1}{\sqrt{2}}v_1 & \frac{v_1v_D}{v_2} \end{array} \right).
\eea
Then, 
the above mass matrix is diagonalized by
\bea
\left(\begin{array}{c} \varphi_1  \\ ({\hat \phi}^0_2)^*  \end{array}\right)  =  \left( \begin{array}{cc} \cos\beta_D & \sin\beta_D \\ \sin\beta_D & -\cos\beta_D  \end{array}\right)  \left( \begin{array}{c} G^+_D \\ {\tilde\varphi}  \end{array}\right). 
\eea
There is one massless complex scalar corresponding to $G^+_D$, and the mass of the dark complex scalar is given by eq.~(\ref{Nmass}).

Finally, we consider the mass matrix for the neutral gauge bosons, $(B^\mu, W^3_\mu, V^0_\mu)$, as follows,
\bea
{\cal M}^2_V &=& \left(\begin{array}{ccc} \frac{1}{4} g^2_Y v^2 & -\frac{1}{4} g g_Y v^2 & \frac{1}{4} g_Y g_D v^2_2 \vspace{0.2cm} \\-\frac{1}{4} g g_Y v^2 & \frac{1}{4} g^2 v^2 & -\frac{1}{4} g g_D v^2_2  \vspace{0.2cm} \\  \frac{1}{4} g_Y g_D v^2_2 &  -\frac{1}{4} g g_D v^2_2 & \frac{1}{4} g^2_D (2v^2_D+v^2_2)  \end{array}\right) \nonumber \\
&=& \left(\begin{array}{ccc}  M^2_Z s^2_W & -M^2_Z c_W s_W & \frac{1}{4} c^{-1}_W e g_D v^2_2\vspace{0.2cm}  \\ -M^2_Z c_W s_W & M^2_Z c^2_W & -\frac{1}{4} s^{-1}_W e g_D v^2_2 \vspace{0.2cm} \\ \frac{1}{4} c^{-1}_W e g_D v^2_2 &  -\frac{1}{4} s^{-1}_W e g_D v^2_2 & m^2_{V^0}    \end{array}\right).
\eea
Then,  the above mass matrix is diagonalized \cite{Bmeson} by
\bea
\left(\begin{array}{c} B_\mu \\ W^3_\mu \\ V^0_\mu  \end{array}\right)  &=&\left( \begin{array}{ccc} c_W & -s_W & 0 \\ s_W & c_W & 0\\ 0 & 0 & 1  \end{array}\right)   \left( \begin{array}{ccc}1 & 0 & 0  \\ 0 & c_\zeta & s_\zeta \\  0 & -s_\zeta & c_\zeta \end{array}\right) \left( \begin{array}{c} A_\mu \\ Z_{1\mu} \\ Z_{2\mu} \end{array}\right)  \nonumber \\
&=&\left( \begin{array}{ccc} c_W & -s_W c_\zeta & -s_W s_\zeta \\ s_W & c_W c_\zeta & c_W s_\zeta   \\ 0 & -s_\zeta & c_\zeta \end{array}\right)   \left( \begin{array}{c} A_\mu \\ Z_{1\mu} \\ Z_{2\mu} \end{array}\right).
\eea
There is one massless gauge boson, which is photon, and the mass eigenvalues for the other neutral gauge bosons \cite{Bmeson} are
\bea
M^2_{Z_{1,2}} = \frac{1}{2} \Big(M^2_Z+m^2_{V^0} \mp \sqrt{(M^2_Z-m^2_{V^0})^2 +4 m^4_{12}} \Big) \label{Zmasses}
\eea
with
\bea
m^2_{12} = -\frac{1}{4} c^{-1}_W s^{-1}_W e g_D v^2_2 =-\frac{s_W g_D}{g_Y}\,M^2_Z\sin^2\beta,
\eea
and the mixing angle between $Z$ and $V^0$ gauge bosons \cite{Bmeson} is given by
\bea
\tan2\zeta= \frac{2m^2_{12} (M^2_{Z_2}-M^2_Z)}{(M^2_{Z_2}-M^2_Z)^2-m^4_{12}}.  \label{Zmix}
\eea

\end{appendix}

\end{document}